%% file: ms.tex
\documentclass[lettersize,journal]{IEEEtran}
\usepackage{orcidlink}
\usepackage{amsmath,amsfonts}
\usepackage{algorithm,setspace}
\usepackage[noend]{algpseudocode}
\usepackage{array}
\usepackage{textcomp}
\usepackage{stfloats}
\usepackage{url}
\usepackage{verbatim}

\usepackage{caption}
\usepackage{subcaption}

\usepackage{graphicx}
\usepackage{cite}
\usepackage{color,soul}
\usepackage{epstopdf}
\usepackage{xfrac}
\usepackage{pifont}
\newcommand{\cmark}{\textcolor{green}{\ding{51}}}%
\newcommand{\xmark}{\textcolor{red}{\ding{55}}}%

\usepackage{mathtools}
\usepackage{cleveref}
\usepackage{csquotes}
\usepackage{booktabs}
\usepackage{multirow}
\usepackage{tabularx,ragged2e}
\newcolumntype{Y}{>{\centering\arraybackslash}X}
\usepackage{footmisc}

\usepackage{varwidth}
\usepackage{amsmath,amssymb,physics}

\usepackage{tikz}
\usepackage{pgf}
\usepackage{pgfplots}
\usepackage{pgfplotstable}
\usetikzlibrary{positioning,backgrounds,fit,dsp,calc,patterns}
\pgfplotsset{compat=1.3}
\usetikzlibrary{shapes.geometric}
\usetikzlibrary{shapes.arrows}
\usepackage{array}
\usepackage[nolist, nohyperlinks]{acronym}
\usepackage{xfrac}

\newcommand{\D}[1]{\mathrm{d}{#1}}
\newcommand{\submin}[0]{_\text{min}}
\newcommand{\submax}[0]{_\text{max}}

\newcommand{\state}[0]{\mathbf{x}_\tau}

\newcommand{\clean}[0]{\mathbf{x}}
\newcommand{\initial}[0]{\mathbf{x}_0}
\newcommand{\final}[0]{\mathbf{x}_T}
\newcommand{\noisy}[0]{\mathbf{y}}
\newcommand{\sco}[0]{\nabla_{\state} \log }
\newcommand{\dnnsco}[0]{\mathbf{s}_\phi}

\tikzstyle{block} = [draw, fill=blue!14, rectangle, minimum height=3em, minimum width=5em, align=center]
\tikzstyle{sum} = [draw, fill=white, circle, node distance=1cm]
\tikzstyle{input} = [coordinate]
\tikzstyle{output} = [coordinate]
\tikzstyle{pinstyle} = [pin edge={to-,thin,black}]
\tikzstyle{branch}=[fill,shape=circle,minimum size=3pt,inner sep=0pt]
\tikzstyle{connarrow}=[-latex, line width=1pt]
\tikzstyle{connline}=[-, line width=1pt]

\begin{document}

\begin{acronym}
\acro{stft}[STFT]{short-time Fourier transform}
\acro{istft}[iSTFT]{inverse short-time Fourier transform}
\acro{dnn}[DNN]{deep neural network}
\acro{pesq}[PESQ]{Perceptual Evaluation of Speech Quality}
\acro{polqa}[POLQA]{perceptual objectve listening quality analysis}
\acro{wpe}[WPE]{weighted prediction error}
\acro{psd}[PSD]{power spectral density}
\acro{rir}[RIR]{room impulse response}
\acro{snr}[SNR]{signal-to-noise ratio}
\acro{lstm}[LSTM]{long short-term memory}
\acro{polqa}[POLQA]{Perceptual Objectve Listening Quality Analysis}
\acro{sdr}[SDR]{signal-to-distortion ratio}
\acro{estoi}[ESTOI]{extended short-term objective intelligibility}
\acro{elr}[ELR]{early-to-late reverberation ratio}
\acro{tcn}[TCN]{temporal convolutional network}
\acro{rls}[RLS]{recursive least squares}
\acro{asr}[ASR]{automatic speech recognition}
\acro{ha}[HA]{hearing aid}
\acro{ci}[CI]{cochlear implant}
\acro{mac}[MAC]{multiply-and-accumulate}
\acro{vae}[VAE]{variational auto-encoder}
\acro{gan}[GAN]{generative adversarial network}
\acro{tf}[T-F]{time-frequency}
\acro{sde}[SDE]{stochastic differential equation}
\acro{drr}[DRR]{direct to reverberant ratio}
\acro{lsd}[LSD]{log spectral distance}
\acro{sisdr}[SI-SDR]{scale-invariant signal to distortion ratio}
\acro{sisir}[SI-SIR]{scale-invariant signal to interference ratio}
\acro{sisar}[SI-SAR]{scale-invariant signal to artifacts ratio}
\acro{mos}[MOS]{mean opinion score}
\acro{wer}[WER]{word error rate}
\acro{map}[MAP]{maximum a-posteriori}
\acro{nfe}[NFE]{number of function evaluations}
\acro{mac}[MAC]{multiply-accumulate}
\acro{rtf}[RTF]{real-time factor}
\end{acronym}

\author{Jean-Marie Lemercier\,{\orcidlink{0000-0002-8704-7658}},~\IEEEmembership{Student Member,~IEEE}, Julius Richter\,{\orcidlink{0000-0002-7870-4839}},~\IEEEmembership{Student Member,~IEEE}, Simon Welker\,{\orcidlink{0000-0002-6349-8462}},~\IEEEmembership{Student Member,~IEEE}, Timo Gerkmann\,{\orcidlink{0000-0002-8678-4699}},~\IEEEmembership{Senior Member,~IEEE}
\thanks{This work has been funded by the Federal Ministry for Economic Affairs and Climate Action, project 01MK20012S, AP380, the German Research Foundation (DFG) in the transregio project Crossmodal Learning (TRR 169) and DASHH (Data Science in Hamburg - HELMHOLTZ Graduate School for the Structure of Matter) with the Grant-No. HIDSS-0002.}%
\thanks{Simon Welker is with the Signal Processing Group, Department of Informatics, Universität Hamburg, 22527 Hamburg Germany, and with the Center for Free-Electron Laser Science, DESY, 22607 Hamburg, Germany (e-mail: simon.welker@uni-hamburg.de).}
\thanks{The other authors are with the Signal Processing Group, Department of Informatics, Universität Hamburg, 22527 Hamburg Germany (e-mail: \{jeanmarie.lemercier; julius.richter; timo.gerkmann\}@uni-hamburg.de).}}

\title{StoRM: A Diffusion-based Stochastic Regeneration Model for Speech Enhancement and Dereverberation}

\maketitle
\begin{abstract}
 Diffusion models have shown a great ability at bridging the performance gap between predictive and generative approaches for speech enhancement. We have shown that they may even outperform their predictive counterparts for non-additive corruption types or when they are evaluated on mismatched conditions. However, diffusion models suffer from a high computational burden, mainly as they require to run a neural network for each reverse diffusion step, whereas predictive approaches only require one pass. As diffusion models are generative approaches they may also produce vocalizing and breathing artifacts in adverse conditions. In comparison, in such difficult scenarios, predictive models typically do not produce such artifacts but tend to distort the target speech instead, thereby degrading the speech quality.
In this work, we present a stochastic regeneration approach where an estimate given by a predictive model is provided as a guide for further diffusion. We show that the proposed approach uses the predictive model to remove the vocalizing and breathing artifacts while producing very high quality samples thanks to the diffusion model, even in adverse conditions. We further show that this approach enables to use lighter sampling schemes with fewer diffusion steps without sacrificing quality, thus lifting the computational burden by an order of magnitude.
Source code and audio examples are available online\footnote{https://uhh.de/inf-sp-storm}.

\end{abstract}


\section{Introduction}
\label{sec:intro}
\input{sections/intro}

\section{Score-based Diffusion Models}
\label{sec:sgmse}
\input{sections/sgmse}

\section{Stochastic Regeneration with Diffusion Models}
\label{sec:refinement}
\input{sections/refinement}

\section{Experimental Setup}
\label{sec:exp}
\input{sections/exp}

\section{Experimental Results and Discussion}
\label{sec:results}

\input{sections/results}

\section{Conclusion}
\label{sec:conclusion}
\input{sections/conclusion}

\bibliographystyle{IEEEtran}
\bibliography{biblio}

\end{document}

%% file: sections/intro.tex
In real-life scenarios and modern communication devices, clean speech sources are often polluted by background noise, interfering speakers, room acoustics and codec degradation~\cite{Naylor2011, Godsill1998DigitalAR}.
We refer to this phenomenon as \textit{speech corruption}, and denote by \textit{speech restoration} the art of recovering clean speech from the corrupted signal~\cite{hendriks2013dft}.
On the one hand, traditional speech restoration methods leverage the statistical properties of the target and interference signals in various domains e.g. time, spectrum, cepstrum or spatial distribution~\cite{gerkmann2018book_chapter}.
On the other hand, machine learning techniques try to learn these statistical properties and how to exploit them from data \cite{wang2018supervised}. Machine learning algortihms can be categorized into predictive (also called discriminative) approaches and generative approaches. We will choose the term \emph{predictive} over \emph{discriminative} as it fits both classification and regression tasks \cite{MurphyBook2}. The field of speech restoration is dominated by predictive approaches that use supervised learning to learn a single best deterministic mapping between corrupted speech $\noisy$ and the corresponding clean speech target  $\clean$ \cite{wang2018supervised}. 
These methods include for instance \ac{tf} masking~\cite{Williamson2017j}, time domain methods~\cite{Luo2018, Lin2021TwoStageBWE} or direct spectro-temporal mapping \cite{Han2017}. They have contributed to drastically increasing the quality of speech restoration algorithms. However, they can distort target speech and suffer from generalizability issues
\cite{Lemercier2022icassp, Richter2022SGMSE++}.

In contrast, generative models implicitly or explicitly learn the target distribution and allow to generate multiple valid estimates instead of a single best estimate as for predictive approaches \cite{MurphyBook2}. Generative approaches include \acp{vae} learning explicit density estimations \cite{kingma2014auto, fang2021variational, leglaive2018variance, richter2020speech}, normalizing flows adding invertible transforms to obtain tractable marginal likelihoods~\cite{Rezende2015Normalizing, Kobyzev2021NFOverview}, \acp{gan} estimating implicit distributions \cite{goodfellow2014generative, Kong2020HiFiGAN} and diffusion approaches \cite{sohl2015deep,ho2020denoising,song2019generative}.
We talk of \textit{conditional} generative models when a covariate $\mathbf c$ is used to guide the generation, leading to the conditional distribution $p(\clean|\mathbf{c})$\cite{MurphyBook2}. This conditioning can either be another modality describing the data (e.g. $\mathbf c$ could be video when $\clean$ is speech), or a modified version of the data, an obvious example being corrupted speech $\noisy$ when the underlying task is speech restoration.
By integrating stochasticity in their latent structure, generative models can capture the inherent uncertainty of the data distribution and produce realistic samples belonging to that distribution rather than a mean of optimal candidates \cite{MurphyBook2}. In doing so, they may obtain better perceptual metrics at the cost of higher point-wise distortion \cite{Whang2021StochasticRefinement}.
In the imaging domain%
, it was observed that predictive approaches tend to brush over the fine-grained details of the considered domain \cite{Whang2021StochasticRefinement, Welker2023}. %
Furthermore, predictive models may result in limited generalization abilities towards unseen noise types or speakers as compared to generative models, which is demonstrated for diffusion-based generative speech enhancement in 
\cite{Richter2022SGMSE++}.

We focus in this work on such diffusion-based generative models, or simply \textit{diffusion models}, which have met great success in generating high-quality samples of natural images \cite{sohl2015deep,ho2020denoising,song2019generative,dhariwal2021diffusion}. 
Diffusion models use a \textit{forward process} to slowly turn data into a tractable prior, usually a standard normal distribution, and train a neural network to solve the \textit{reverse process} to generate clean data from this prior~\cite{Yang2022DMOverview}.
These diffusion models can also be used for conditional generation in restoration tasks, which has recently been proposed for speech processing tasks such as enhancement and dereverberation \cite{Welker2022SGMSE, Richter2022SGMSE++, lu2022conditional, serra2022universal} as well as bandwidth extension \cite{Lemercier2022icassp, Han2022NUWave2}. 

One limiting aspect of diffusion models is their heavy computational burden. Several steps are needed for reverse diffusion, each of them calling the neural network used for score estimation. Much effort has been recently put into reducing this number of steps, either by optimization of the reverse noise schedule \cite{Kingma2021VDM}, modifications in the formulation of the diffusion processes \cite{Song2021DDIM, Lam2022BDDM}, or projection into a latent space \cite{ROmbach2022StableDiffusion} or a reduced subspace \cite{Jing2022Subspace}.
We also observed in past experiments that our previously proposed diffusion model is prone to confuse phonemes and generate vocalizing artifacts when facing very adverse conditions. This is due to the generative behaviour of the model under high uncertainty over the presence or nature of speech, and this naturally leads to a degradation, e.g. in \ac{asr}.

In this work, we propose a \textit{stochastic regeneration} scheme combining predictive and generative models to produce high quality samples while reducing the computational burden of diffusion models and their tendency to generate unwanted artifacts. We propose to first use a predictive approach to estimate a restored version of the corrupted speech. This estimate is then used as a guide by a diffusion model, which requires only a few diffusion steps to output a final clean speech estimation where the distortions introduced by the predictive stage are corrected while vocalizing artifacts and phonetic confusions are avoided. Both listening experiments and instrumental metrics confirm an impressive state-of-the-art perceptual quality of our proposed approach.
Other refinement approaches using diffusion models were recently proposed. The \textit{stochastic refinement} approach \cite{Whang2021StochasticRefinement, Qiu2023SRTNet} subtracts the output of the predictive model from the corrupted speech, and this residual is used for further estimation by a diffusion model. We argue hereafter that learning the residual is however a hard task and demonstrate that our approach outperforms this stochastic refinement in terms of instrumentally measured speech quality. Another refinement approach using diffusion models is \textit{denoising diffusion restoration models} \cite{kawar2022denoising, Saito2022DerevRefiner, Sawata2022refiner}, where the corruption operator is assumed to be known (or at least its singular value decomposition) and is used to modify the reverse diffusion process at inference time.

We evaluate our proposed approach for speech enhancement with low input \acp{snr} and speech dereverberation, using clean speech from the WSJ0 corpus \cite{datasetWSJ0}. We also show \ac{asr} results on the TIMIT dataset \cite{Garofolo1992TIMIT}, and report results on the standardized Voicebank/DEMAND dataset \cite{valentini2016investigating}. Ablation studies are performed on sampling efficiency, intial predictor mismatch and training strategy.

%% file: sections/sgmse.tex
Diffusion models originally use discrete-time diffusion processes modeled by Markov chains \cite{ho2020denoising}. They have been recently extended to continuous-time diffusion processes formulated by \acp{sde} in \cite{song2021sde}, allowing for new training paradigms such as score matching \cite{hyvarinen2005estimation, vincent2011connection}. This class model is subsequently denoted as \textit{score-based diffusion models}.
Score-based diffusion models are defined by three  components: a forward diffusion process, a score function estimator, and a sampling method for inference.

\subsection{Forward and reverse processes}\label{sec:processes}

The stochastic forward process $\{\state\}_{\tau=0}^T$ used in score-based diffusion models is defined as an Itô \ac{sde} \cite{Oksendal2000SDE, song2021sde}:
    \begin{equation} \label{eq:forward-sde}
        \D{\state} = \mathbf{f}(\state, \tau) \D{\tau} + g(\tau) \D{\mathbf w},
    \end{equation}
where $\state$ is the current state of the process indexed by $\tau \in [0, T]$ with the initial condition representing clean speech $\mathbf{x}_0 = \mathbf{x}$. The continuous \textit{diffusion time} variable $\tau$ relates to the progress of the stochastic process and should not be mistaken for our usual notion of \textit{signal time}. As our process is defined in the complex spectrogram domain, independently for each \ac{tf} bin, the variables in bold are vectors in $\mathbb C^d$ containing the coefficients of a flattened complex spectrogram--- with $d$ the product of the time and frequency dimensions--- whereas variables in regular font represent real scalar values. 
The set $\{\state\}_{\tau \in ]0, T[}$ can be seen as latent variables used to parameterize the conditional distribution $p(\state|\mathbf{x}_0, \noisy)$.
The stochastic process $\mathbf w$ denotes a standard $d$-dimensional Brownian motion, that is, $\D{\mathbf w}$ is a zero-mean Gaussian variable with standard deviation $\D{\tau}$ for each \ac{tf} bin.

The \emph{drift} function $\mathbf f$ and \emph{diffusion} coefficient $g$ as well as the initial condition $\initial$ and the final diffusion time $T$ uniquely define the Itô process $\{\state\}_{\tau=0}^T$\cite{Oksendal2000SDE}.
Under some regularity conditions on $\mathbf f, g$ allowing a unique and smooth solution to the Kolmogorov equations associated to \eqref{eq:forward-sde}, the reverse process $\{\state\}_{\tau=T}^0$ is another diffusion process defined as the solution of the following \ac{sde} \cite{anderson1982reverse, song2021sde}:
\begin{equation}\label{eq:reverse-sde}
        \resizebox{0.9\hsize}{!}{%
        $\D{\state} = \left[
            -\mathbf f(\state, \tau) + g(\tau)^2\sco  p_\tau(\state)
        \right] \D{\tau}
        + g(\tau)\D{\bar{\mathbf w}},$
        }
\end{equation}
where $\D{\bar{\mathbf w}}$ is a $d$-dimensional Brownian motion for the time flowing in reverse and $\sco p_\tau(\state)$ is the \emph{score function}, i.e. the gradient of the logarithm data distribution for the current process state $\state$.

Speech restoration tasks can be regarded either as one-to-one mapping tasks between corrupted speech $\noisy$ and $\mathbf{x}_0$, which leads to predictive modelling; or as conditional generation tasks, i.e. generation of  $\mathbf{x}_0$ conditioned on $\noisy$.
Previous diffusion-based approaches proposed to condition the process explicitly within the neural network \cite{chen2021wavegrad} or through guided classification \cite{dhariwal2021diffusion}. 
In \cite{Welker2022SGMSE}, the conditioning is directly incorporated into the diffusion process by defining the forward process as the solution to the following \ac{sde}:
\begin{equation}\label{eq:ouve-sde}
        \resizebox{0.9\hsize}{!}{%
    $\D{\state} = \underbrace{\gamma(\noisy-\state)}_{:=\,\mathbf f(\state, \noisy)} \D{\tau}
        + \underbrace{\left[ \sigma\submin \left(\frac{\sigma\submax}{\sigma\submin}\right)^\tau \sqrt{2\log\left(\frac{\sigma\submax}{\sigma\submin}\right)} \right]}_{:=\,g(\tau)} \D{\mathbf w}$.
        }
\end{equation}
This equation belongs to the class of Ornstein-Uhlenbeck \acp{sde} \cite{Oksendal2000SDE}, a subclass of Itô \acp{sde} in which the drift function $\mathbf f$ is affine in $\state$ and does not depend on $\tau$, and the diffusion coefficient $g$ only depends on $\tau$. The equation introduces a \emph{stiffness} hyperparameter $\gamma$ controlling the slope of the decay from $\noisy$ to $\initial$, and $\sigma\submin$ and $\sigma\submax$ are two hyperparameters controlling the \emph{noise scheduling}, that is, the amount of Gaussian white noise injected at each timestep of the process. 

The interpretation of our forward process in Eq. \eqref{eq:ouve-sde}, visualized on Fig.~\ref{fig:process}, is as follows: at each time step and for each \ac{tf} bin independently, an infinitesimal amount of corruption is added to the current process state $\state$, along with Gaussian noise with standard deviation $g(\tau) \D{\tau}$. 
Therefore, the mean of the current process decays exponentially towards $\noisy$ while the variance increases as in the variance-exploding scheme of Song et al. \cite{song2021sde}, leading to a final distribution $\state$ which is the corrupted signal $\noisy$ with some additional Gaussian noise.
Given an initial condition $\initial$ and the covariate $\noisy$, the solution to \eqref{eq:ouve-sde} admits the following complex Gaussian distribution for the process state $\state$ called \emph{perturbation kernel}:
\begin{equation}
\label{eq:perturbation-kernel}
    p_{0, \tau}(\state|\initial, \noisy) = \mathcal{N}_\mathbb{C}\left(\state; \boldsymbol \mu(\initial, \noisy, \tau), \sigma(\tau)^2 \mathbf{I}\right),
\end{equation}
Following \cite{sarkka2019sde}, we determine closed-form solutions for the mean $\boldsymbol \mu$ and variance $\sigma(\tau)^2$:
\begin{equation}
\label{eq:mean}
    \boldsymbol\mu(\initial,\noisy, \tau) = \mathrm e^{-\gamma \tau} \initial + (1-\mathrm e^{-\gamma \tau}) \noisy
    \,,
\end{equation}%
\begin{equation}
    \label{eq:std}
    \sigma(\tau)^2 = \frac{
        \sigma\submin^2\left(\left(\sfrac{\sigma\submax}{\sigma\submin}\right)^{2\tau} - \mathrm e^{-2\gamma \tau}\right)\log(\sfrac{\sigma\submax}{\sigma\submin})
    }{\gamma+\log(\sfrac{\sigma\submax}{\sigma\submin})}
    \,.
\end{equation}
\vspace{-1em}

\input{plots/process.tex}

\subsection{Score function estimator}\label{sec:training}

When performing inference, one tries to solve the reverse \ac{sde} in Eq. \eqref{eq:reverse-sde}. In the general case, the score function $\sco p_\tau(\state)$ is not readily available, it can however be estimated by a \ac{dnn} $\dnnsco$ called the \emph{score model}. Given the simple Gaussian form of the perturbation kernel $p_{0, \tau}(\state|\initial, \noisy)$ \eqref{eq:perturbation-kernel} and the regularity conditions exhibited by the mean and variance, a \emph{denoising score matching} objective can be used to train the score model $\dnnsco$ \cite{hyvarinen2005estimation, vincent2011connection}.

The score function of the perturbation kernel is:
\begin{equation}
    \sco p_{0, \tau}(\state|\initial, \noisy) = -\frac{\state - \boldsymbol\mu(\initial,\noisy, \tau)}{\sigma(\tau)^2}.
\end{equation}
Once a clean utterance $\mathbf{x}_0$ and noisy utterance $\noisy$ are picked in the training set, the current process state is obtained as $\state =  \boldsymbol\mu(\initial,\noisy, \tau) + \sigma(\tau) \mathbf z$,  with $\mathbf z \sim \mathcal{N}_\mathbb{C}\left(\mathbf z; \mathbf 0, \mathbf{I}\right)$. We can therefore write the denoising score matching objective as follows \cite{song2021sde}:
\begin{equation}\label{eq:training-loss}
    \mathcal{J}^{(\mathrm{DSM})}(\phi)
    = \mathbb{E}_{t,(\initial,\noisy), \mathbf z, \state} \left[
        \norm{\mathbf s_\phi(\state, \noisy, \tau) + \frac{\mathbf z}{\sigma(\tau)}}_2^2
    \right].
\end{equation}
Here, we sample $\tau$ sampled uniformly in $[\tau_\epsilon, T]$ where $\tau_\epsilon$ is a minimal diffusion time used to avoid numerical instabilities.
This approach is analogous to the denoising objective used in the discrete-time formulation by \cite{ho2020denoising}, where one estimates the noise added at each step to learn the reverse process.

\subsection{Inference through reverse sampling}\label{sec:inference}

At inference time, we first sample an initial condition of the reverse process, corresponding to $\final$, with:
\begin{equation}
    \final \sim \mathcal N_{\mathbb C}(\final; \noisy, \sigma^2(T) \mathbf I), 
\end{equation}
This sample only approximates the training condition, as the final process distribution $p_{T}(\final)$ does not perfectly match $p(\noisy)$ (see \figurename~\ref{fig:process}).

Conditional generation is then performed by solving the so-called \emph{plug-in reverse \ac{sde}} from $\tau=T$ to $\tau=0$, where the score function is replaced by its estimator $\mathbf s_\phi$, assuming the latter was trained e.g. according to Section~\ref{sec:training}:
\begin{equation}\label{eq:plug-in-reverse-sde}
    \D{\state} =
        \left[
            -\mathbf f(\state, \noisy) + g(\tau)^2\mathbf s_\phi(\state, \noisy, \tau)
        \right] \D{\tau}
        + g(\tau)\D{\bar{\mathbf w}}
\end{equation}

We use classical numerical solvers based on a discretization of 
\eqref{eq:plug-in-reverse-sde} according to a uniform grid of $N$ points on the interval $[0, T]$ (no minimal diffusion time is needed here). Classical solvers include the Euler-Maruyama method, higher-order single-step methods, and predictor-corrector sampler schemes \cite{song2021sde}. In the latter, at each reverse step $\tau$, the predictor uses a single-step method like Euler-Maruyama to generate $\state$, and the corrector uses the output of the score network to ensure consistency of the resulting sample with a marginal distribution consistent with the score estimate.

For notational convenience, we will denote by $G_\phi$ the generative model corresponding to the reverse diffusion process solver parameterized by the plug-in \ac{sde} \eqref{eq:plug-in-reverse-sde} and the score network $\dnnsco$, such that the final estimate is $\widehat{\mathbf{x}} = G_\phi(\noisy)$.

%% file: plots/process.tex
\begin{figure*}[t]
    \caption{\centering\textit{Visualization of the forward and backward processes in \eqref{eq:ouve-sde}. Mean curve \eqref{eq:mean} is in solid black and variance \eqref{eq:std} is represented by the greyed area. Several realizations of the diffusion process are represented by thin black lines. The mismatch between $p_\tau$ centered on $\mathbf{x_\tau}$ and $\tilde{p}_\tau$ centered on $\mathbf{y}$ comes from the fact that the mean in \eqref{eq:mean} can not reach $\mathbf{y}$ in finite time. This mismatch causes unavoidable bias in the reverse process, even were the score perfectly known.}}
    \centering
    \includegraphics[width=\textwidth]{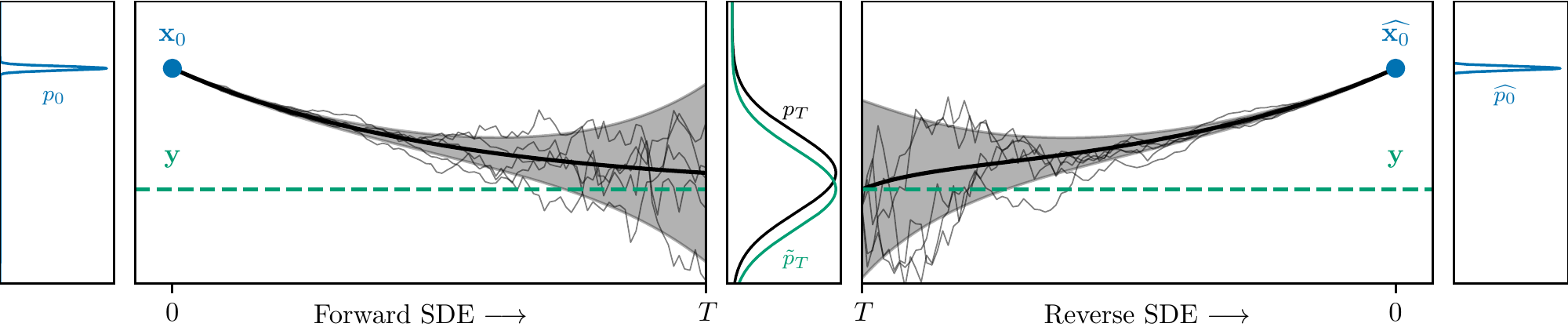}
    \label{fig:process}
\vspace{-1em}
\end{figure*}

%% file: sections/refinement.tex
\subsection{Predictive artifacts for images and spectrograms}

Most problems in speech restoration (e.g. denoising without knowledge of the environmental noise signal, dereverberation or bandwidth extension) are ill-posed inverse problems. This means that either (i) it is not possible to exactly retrieve $\clean$ given $\noisy$ (e.g. dereverberation with a known non-minimum-phase room impulse response); or (ii) many versions of clean speech $\clean^{(i)}$ can correspond to the same corrupted speech $\noisy$ (e.g. bandwidth extension).
Consider a predictive model $D_\theta$ trained with a $L^2$ regression objective $\mathbb{E}_{(\clean, \noisy)} || \clean - D_\theta(\noisy)||_2^2$. Because of the expectation over all training examples, an optimal predictive model learns the mapping to the posterior mean $\noisy \rightarrow \mathbb{E}[\clean|\noisy]$, thereby minimizing the \textit{average} distortion over all training examples.
This phenomenon is known as \textit{regression to the mean} \cite{Whang2021StochasticRefinement, MurphyBook2}. This can be problematic if the posterior distribution $p(\clean | \noisy)$ has an intricate structure and is thus not well represented by its mean $\mathbb{E}[\clean | \noisy]$ (see Figure~\ref{fig:clouds} here, or Figure~1 in \cite{delbracio2023inversion}).

In image processing problems, this translates to predictive approaches being incapable of reproducing fine-grained details like e.g. edges and hair structure in human portraits \cite{Whang2021StochasticRefinement, Welker2023}. Our interpretation is that, given that these regions have the highest variability across natural image data \cite{Kendall2017}, mapping directly to the posterior mean $\mathbb{E}[\clean | \noisy]$ will smooth out these details by mistaking them for noise. Therefore the predictive model will output a sample which does not necessarily lie on the posterior data manifold.
When training predictive models for spectrogram estimation, we also observe that the models tend to introduce distortions in the target speech when the corruption level is high, leading to \textit{overdenoising} effects and loss of output resolution \cite{Lemercier2022icassp, Avila2018}.

\input{plots/tikz_comparison_img_spec}
The link between these observations in the imaging and speech domains is the following: performing speech restoration in the spectrogram domain can be seen as image processing (with two pixel dimensions representing the real and imaginary parts instead of three for RGB image processing). The distortions observed in the predictive model output are a smoothing effect \textit{in the spectrogram seen as an image}. This results in removal of fine-grained detail corresponding to quiet regions (i.e. low luminosity detail in images) and onsets or offsets (i.e. edges in images) in the spectrogram.
This is visualized in the third column of \figurename~\ref{fig:comparison_img_vs_spectrogram}, where we directly compare spectrograms from our previous study \cite{Lemercier2022icassp} with images in Welker et al.~\cite[\figurename~1]{Welker2023} reproduced here.

Several paradigms using generative modelling can be envisaged to correct this bias of the predictive model without having to resort to a full-fletched computationally heavy diffusion-based generative model.
Next, we present two of these approaches, namely \textit{stochastic refinement} by Whang et al.~\cite{Whang2021StochasticRefinement} and \textit{stochastic regeneration} which we propose here.

\subsection{Stochastic refinement}
\input{sections/enhance_and_refine}

\subsection{Stochastic regeneration}

For \textit{stochastic regeneration} we propose to cascade the predictive model $D_\theta$ and the generative diffusion model $G_\phi$. The generative model learns to \textit{regenerate} the clean speech based on the distorted version provided by the predictive approach. This is conceptually different from the stochastic refinement approach, where the target cues exist in the residual (but are hard to access given the noise present in the noisy residue $\mathbf{r}$) and need to be \textit{refined} by the diffusion model.

The task of the diffusion model is then to guide generation of the clean speech $\initial$ given the first estimate $D_\theta(\noisy)$. If we look at the decomposition in \eqref{eq:decompo}, we simply have to remove the residual noise $\tilde{\mathbf{n}}$ and restore the distorted target cues $\mathbf{x^{(dis)}}$. The resulting \textit{a priori} \ac{snr} in the starting point (again without considering the added Gaussian noise) is very high, as for a reasonable predictor  $||\clean^{(\mathrm{dis})}||, ||\tilde{\mathbf{n}}|| \ll ||\mathbf{n}||$.
The estimate is then obtained as:
\begin{equation}
    \hat{\clean} = G_\phi( D_\theta( \noisy ) )
\end{equation}

The inference process is shown in \figurename~\ref{fig:stochastic-regeneration-inference}.
We name the resulting \textbf{Sto}chastic \textbf{R}egeneration \textbf{M}odel \textit{StoRM}.

For training, we use a criterion $\mathcal{J}^{(\mathrm{StoRM})}$ combining denoising score matching 
and a supervised regularization term 
---e.g. mean square error---matching the output of the initial predictor to the target speech:
\begin{equation} \label{eq:regen-loss}
\begin{aligned}
    &\mathcal{J}^{(\mathrm{DSMS})}(\phi) = \mathbb{E}_{\tau,(\clean,\noisy), \mathbf z} 
        \norm{\mathbf \dnnsco(\state, \left[ \noisy, D_\theta(\noisy) \right],  \tau) + \frac{\mathbf z}{\sigma(\tau)}}_2^2,  \\[1pt]
    &\mathcal{J}^{(\mathrm{Sup})}(\theta) =
    \mathbb{E}_{(\clean,\noisy)} 
        \norm{\clean - D_\theta(\noisy) }_2^2,  \vspace{1em} \\[6pt]
    &\mathcal{J}^{(\mathrm{StoRM})}(\phi, \theta) = \mathcal{J}^{(\mathrm{DSMS})}(\phi) + \alpha \mathcal{J}^{(\mathrm{Sup})}(\theta),
    \end{aligned}
\end{equation}

\noindent where $\alpha$ is a balance term that we empirically set to $1$.

One may object that the estimate $D_\theta(\noisy)$ is not a sufficient statistic for the model to reconstruct target cues. However, by their very learning principle, generative models are able to create data based on the clean examples seen during training, hence our choice of the terminology \textit{stochastic regeneration}. 
Stochastic regeneration is still a generative model with respect to the definition given in introduction, as it is able to output realistic samples belonging to a posterior distribution. We visually summarize the concepts of predictive, generative and our model in \figurename~\ref{fig:clouds}.
We describe in algorithms \ref{alg:training} and \ref{alg:inference} the training and inference stages of StoRM, respectively.

\input{plots/tikz_stochastic_regeneration_inference}

\input{plots/algo_training}

\input{plots/algo_inference}

\input{plots/clouds}

%% file: plots/tikz_comparison_img_spec.tex
\newcommand{\spectrow}{.31\textwidth}
\newcommand{\spectroxs}{-.04\textwidth}
\newcommand{\spectroys}{-.185\textwidth}
\newcommand{\xmax}{6}

\begin{figure*}
\caption{\centering \textit{Visualization of samples obtained with predictive approach (NCSN++M, see Section \ref{sec:exp}) and generative model (SGMSE+M, see \cite{Richter2022SGMSE++} and Section \ref{sec:exp}) for two ill-posed problems, namely speech dereverberation (top, from \cite{Lemercier2022icassp}) and JPEG artifact removal (bottom, from \cite{Welker2023}). Spectrograms horizontal and vertical axes represent time and frequency respectively.}}
\centering

\begin{tikzpicture}[scale=0.98, transform shape]
 
\begin{axis}
[   title style={yshift=-2.6ex},
    name={clean},
    title = {Clean},
    hide axis,
    width =\spectrow,
    height =\spectrow
]
\addplot graphics[xmin=0,ymin=0,xmax=\xmax,ymax=8000] {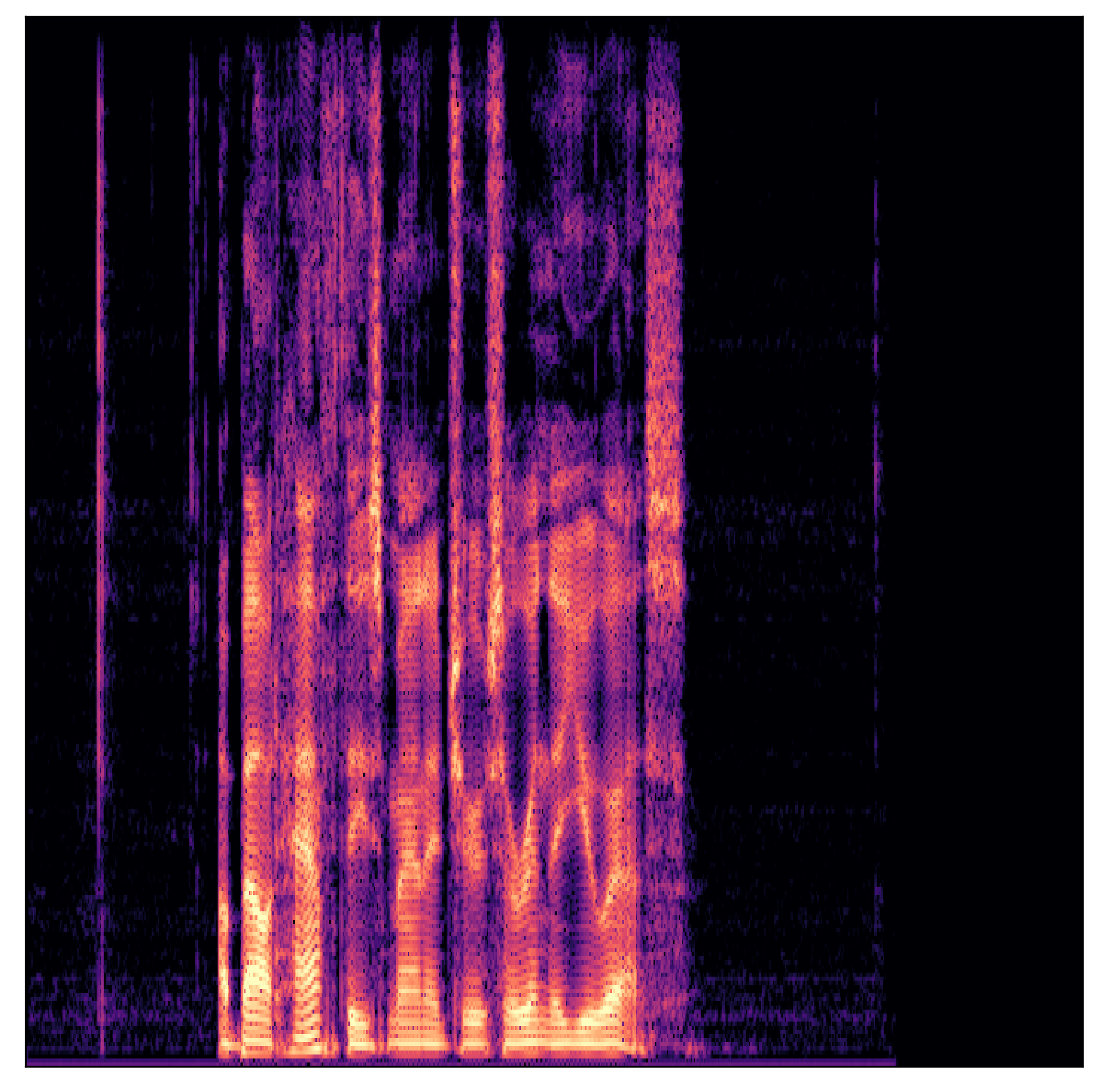};
\end{axis}

\begin{axis}
[   title style={yshift=-2.6ex},
    name={noisy},
    title = {Noisy},
    at = {(clean.south east)},
    hide axis,
    xshift = \spectroxs,
    width =\spectrow,
    height =\spectrow
]
\addplot graphics[xmin=0,ymin=0,xmax=\xmax,ymax=8000] {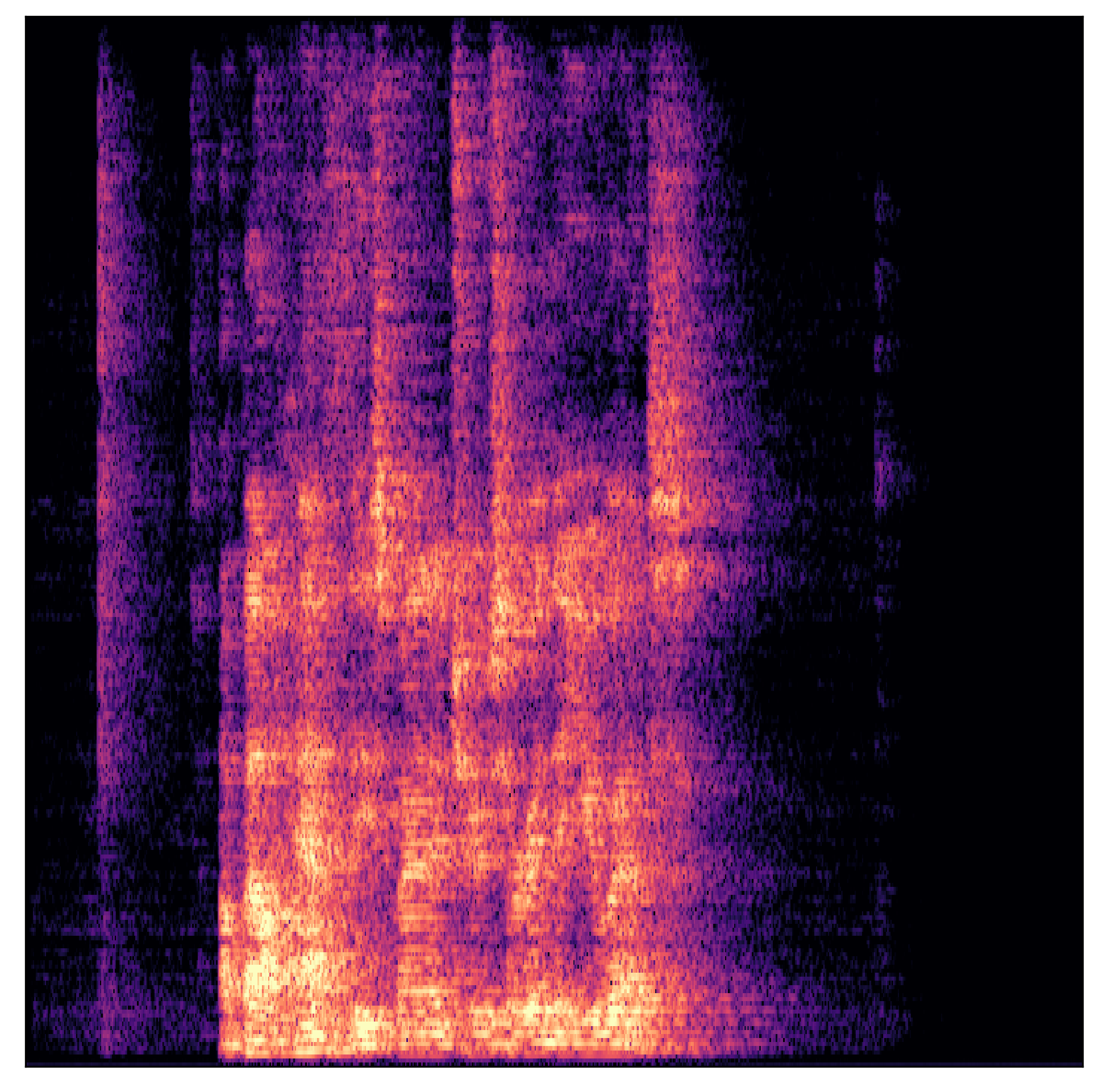};
\end{axis}
 
\begin{axis}
[   title style={yshift=-2.6ex},
    name={predictor},
    title = {Predictive},
    at = {(noisy.south east)},
    hide axis,
    xshift = \spectroxs,
    width =\spectrow,
    height =\spectrow
]
\addplot graphics[xmin=0,ymin=0,xmax=\xmax,ymax=8000] {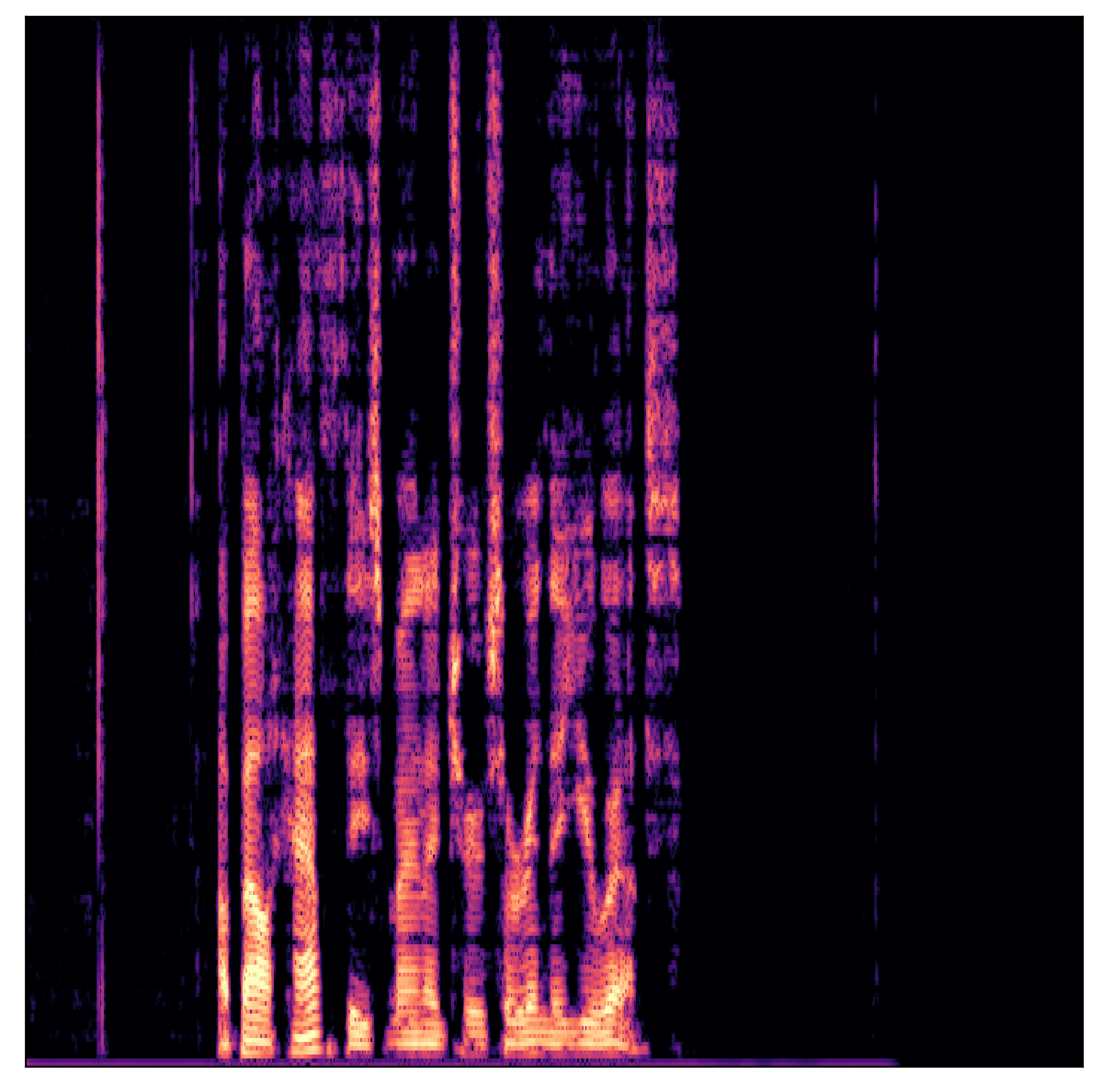};
\end{axis}
 
\begin{axis}
[   title style={yshift=-2.6ex},
    name={residual},
    title = {Generative},
    at = {(predictor.south east)},
    hide axis,
    xshift = \spectroxs,
    width =\spectrow,
    height =\spectrow
]
\addplot graphics[xmin=0,ymin=0,xmax=\xmax,ymax=8000] {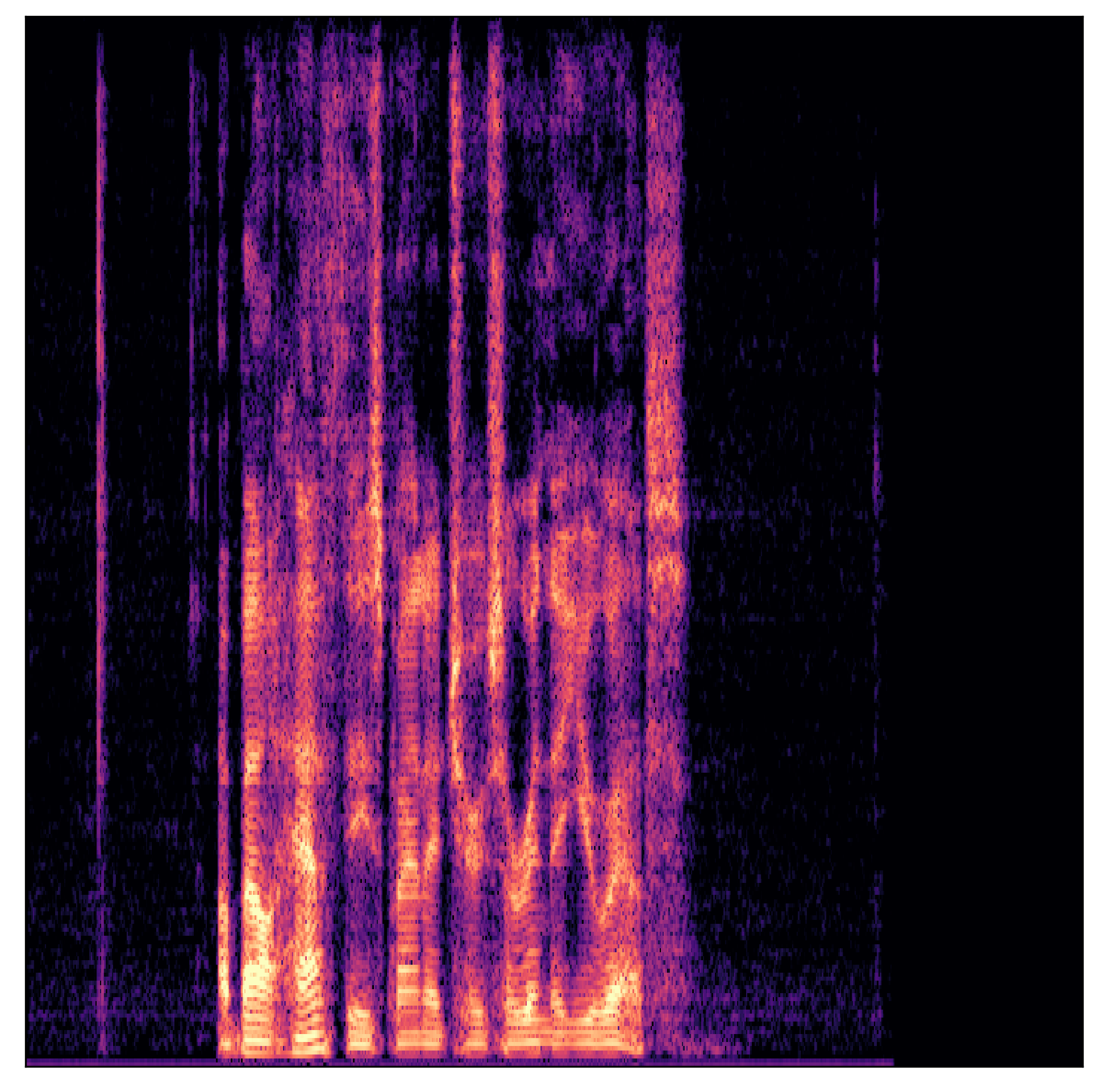};
\end{axis}

\begin{axis}
[
    name={clean2},
    hide axis,
    at = {(clean.south west)},
    yshift = -.1776\textwidth,
    xshift = 0.0048\textwidth,
    width = .3015\textwidth,
    height = .3015\textwidth,
]
\addplot graphics[xmin=0,ymin=0,xmax=1,ymax=1] {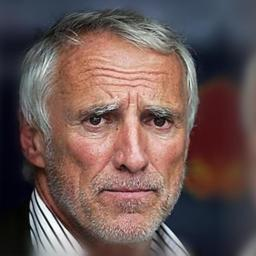};
\end{axis}

\begin{axis}
[
    name={noisy2},
    title = {},
    at = {(clean2.south east)},
    hide axis,
    xshift = -.03125\textwidth,
    width = .3015\textwidth,
    height = .3015\textwidth,
]
\addplot graphics[xmin=0,ymin=0,xmax=1,ymax=1] {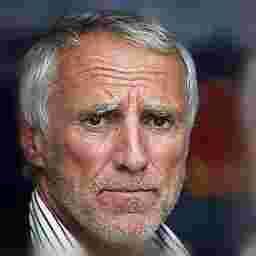};
\end{axis}
 
\begin{axis}
[
    name={predictor2},
    title = {},
    at = {(noisy2.south east)},
    hide axis,
    xshift = -.03125\textwidth,
    width = .3015\textwidth,
    height = .3015\textwidth,
]
\addplot graphics[xmin=0,ymin=0,xmax=1,ymax=1] {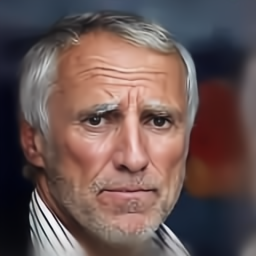};
\end{axis}
 
\begin{axis}
[
    name={residual2},
    title = {},
    at = {(predictor2.south east)},
    hide axis,
    xshift = -.03125\textwidth,
    width = .3015\textwidth,
    height = .3015\textwidth,
]
\addplot graphics[xmin=0,ymin=0,xmax=1,ymax=1] {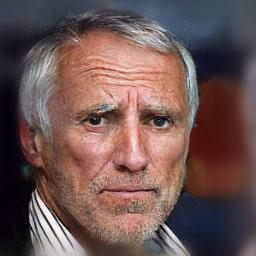};
\end{axis}

\end{tikzpicture}

\vspace{-1em}
\label{fig:comparison_img_vs_spectrogram}
\end{figure*}

%% file: sections/enhance_and_refine.tex
Instead of solving the reverse diffusion process from noisy speech to a clean speech estimate, the \textit{stochastic refinement} approach by Whang et al. \cite{Whang2021StochasticRefinement} uses both a predictive approach and a generative diffusion model for efficient inference. 

A predictive model $D_\theta$ serves as an \textit{initial predictor} producing an estimate $D_\theta(\noisy)$. This estimate often lacks fine-grained detail and has significant target speech distortions, especially for corruption models like reverberation \cite{Lemercier2022icassp}.
Let us write the predictive model output as:
\begin{equation}\label{eq:decompo}
    D_\theta(\noisy) = \clean - \clean^{(\mathrm{dis})} + \tilde{\mathbf{n}}.
\end{equation}
The \textit{target distortion} $\clean^{(\mathrm{dis})}$ is the artifact introduced by the predictive model: it contains target cues that were mistaken for corruption by the model, and consequently distorted. The residual corruption $\tilde{\mathbf{n}}$ is what remains of the interference (e.g. noise or reverberation) after being processed by the model. There is behind this decomposition an underlying orthogonality assumption between $\clean$ and $\mathbf{n}$, which implies orthogonality between $\clean^{(\mathrm{dis})}$ and $\tilde{\mathbf{n}}$ \cite{Vincent2006}.

A diffusion-based generative model $G_\phi$ is then used to learn the distribution of the \textit{ideal residue} $\mathbf{r_x} = \clean - D_\theta(\noisy)$, starting from the \textit{noisy residue} $\mathbf{r_y} = \noisy - D_\theta(\noisy)$. 
Finally, the ideal residue estimate is added to the predictor estimate:
\begin{align}
    \mathbf{\widehat{x}} &= D_\theta(\noisy) + \mathbf{\widehat{r_x}} \\
     &= D_\theta(\noisy) + G_\phi( \noisy - D_\theta ( \noisy ))
\end{align}

\input{plots/residuals.tex}
Results in \cite{Whang2021StochasticRefinement, Qiu2023SRTNet} seem to indicate that this stochastic refinement approach performs as expected, outperforming the initial predictor on perceptual metrics and the pure generative approach with fewer diffusion steps. However, we argue that learning the residual is suboptimal as the residual data distribution $p(\mathbf{r_x})$ does not have a structure like the target data distribution $p(\clean)$. 
Indeed, using \eqref{eq:decompo}, one can rewrite $\mathbf{r_x}$ as:
\begin{align}
    \mathbf{r_x} &= \clean - D_\theta(\noisy) \nonumber \\
    &= \clean^{(\mathrm{dis})} - \tilde{\mathbf{n}},
\end{align}
and notice that the distribution of $\mathbf{r_x}$ highly depends on the choice of the predictive model as well as on the task, which does not assure a structured distribution in the general case. 
We show examples in \figurename~\ref{fig:residuals} of residuals generated by predictive models for speech enhancement and dereverberation, which confirm this observation. 
For dereverberation (similarly for deblurring, shown in \cite{Whang2021StochasticRefinement}), the residue has an overall structure somewhat similar to the target, because of the convolutional corruption model.
However, the formants structure is severely degraded.
For denoising, the residue has no clear structure (compared to e.g. clean speech) and we thus argue that it cannot be easily estimated by a generative process.
In Whang et al.~\cite{Whang2021StochasticRefinement}, it is shown that the residual distribution has lower entropy per pixel than the original distribution, which makes learning the residual easier. This is also true for our denoising and dereverberation tasks. However the pointwise entropy of a distribution relates to the quantity of information that needs to be learnt by the model and does not capture global structures in the data which can actually help facilitate training.
Most importantly, when rewriting the noisy residue $\mathbf{r_y}$ as:
\begin{align}
    \mathbf{r_y} &= \noisy - D_\theta(\noisy) \nonumber \\
    &= \clean^{(\mathrm{dis})} + \mathbf{n} - \tilde{\mathbf{n}},
\end{align}
\noindent one notices that the resulting \textit{a priori} \ac{snr} of the starting point of the reverse process (without accounting for the added Gaussian noise) is very low, as $|| \clean^{(\mathrm{dis})}||, ||\tilde{\mathbf{n}}|| \ll ||\mathbf{n}||$ for low-enough \acp{snr} and good-enough initial predictor.
This makes learning difficult, and we therefore propose to use a different refinement process that we denote as \textit{stochastic regeneration}.

%% file: plots/residuals.tex
\renewcommand{\spectrow}{.28\textwidth}
\renewcommand{\spectroxs}{.00\textwidth}
\renewcommand{\spectroys}{-.193\textwidth}
\renewcommand{\xmax}{5}
\begin{figure*}

\caption{\protect\centering \textit{Log-energy spectrograms of clean, noisy, processed and residual utterances for denoising (top) and dereverberation (bottom). The predictor used is NCSN++M
.}}
\centering
\begin{tikzpicture}[scale=0.95, transform shape]
 
\begin{axis}
[   title style={yshift=-1.0ex},
    name={clean},
    title = {$\mathbf{x}$},
    axis line style={draw=none},
    xmin = 0, xmax = \xmax,
    ymin = 0, ymax = 8000,
    yticklabels = {0,0,2,4,6},
    yticklabel style = {yshift=5pt},
    xticklabels=\empty,
    ylabel = {Frequency [Hz]},
    width =\spectrow,
    height =\spectrow
]
\addplot graphics[xmin=0,ymin=0,xmax=\xmax,ymax=8000] {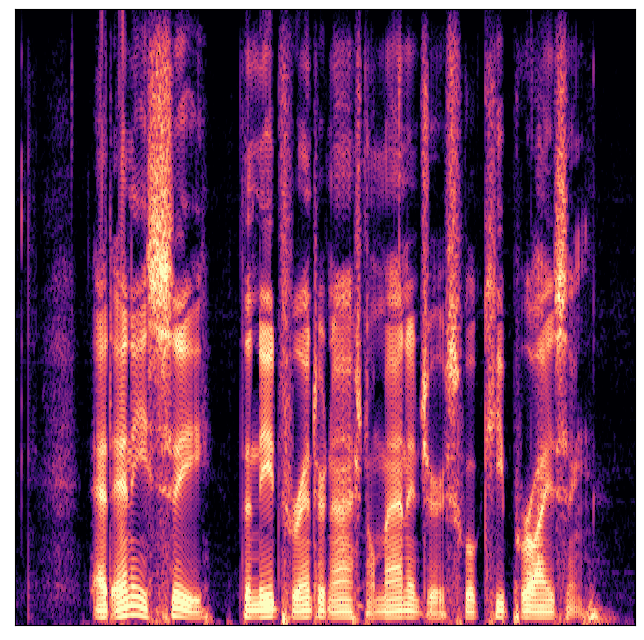};
\end{axis}

\begin{axis}
[   title style={yshift=-1.42ex},
    name={noisy},
    title = {$\mathbf{y}$},
    at = {(clean.south east)},
    xshift = \spectroxs,
    axis line style={draw=none},
    xmin = 0, xmax = \xmax,
    ymin = 0, ymax = 8000,
    xticklabels=\empty,
    yticklabels=\empty,
    width =\spectrow,
    height =\spectrow
]
\addplot graphics[xmin=0,ymin=0,xmax=\xmax,ymax=8000] {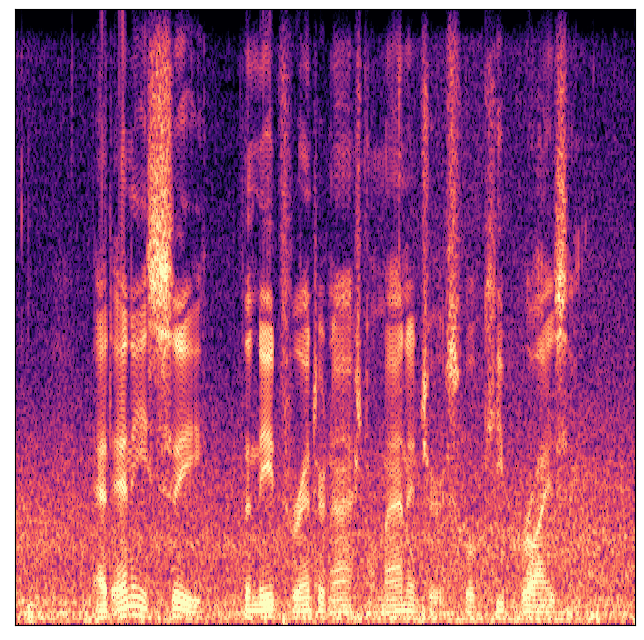};
\end{axis}
 
\begin{axis}
[   title style={yshift=-1.6ex},
    name={predictor},
    title = {$D_\theta(\mathbf{y})$},
    at = {(noisy.south east)},
    xshift = \spectroxs,
    axis line style={draw=none},
    xmin = 0, xmax = \xmax,
    ymin = 0, ymax = 8000,
    xticklabels=\empty,
    yticklabels=\empty,
    width =\spectrow,
    height =\spectrow
]
\addplot graphics[xmin=0,ymin=0,xmax=\xmax,ymax=8000] {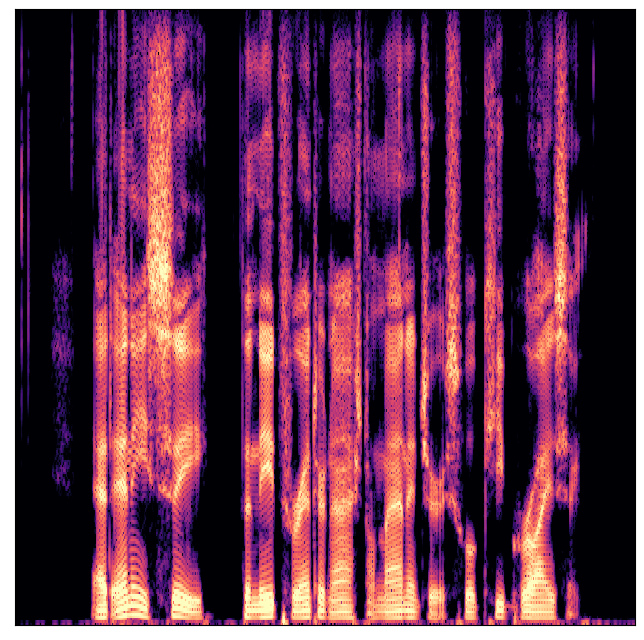};
\end{axis}
 
\begin{axis}
[   title style={yshift=-1.6ex},
    name={residual},
    title = {$\mathbf{x} - D_\theta(\mathbf{y})$},
    at = {(predictor.south east)},
    xshift = \spectroxs,
    axis line style={draw=none},
    xmin = 0, xmax = \xmax,
    ymin = 0, ymax = 8000,
    xticklabels=\empty,
    yticklabels=\empty,
    width =\spectrow,
    height =\spectrow
]
\addplot graphics[xmin=0,ymin=0,xmax=\xmax,ymax=8000] {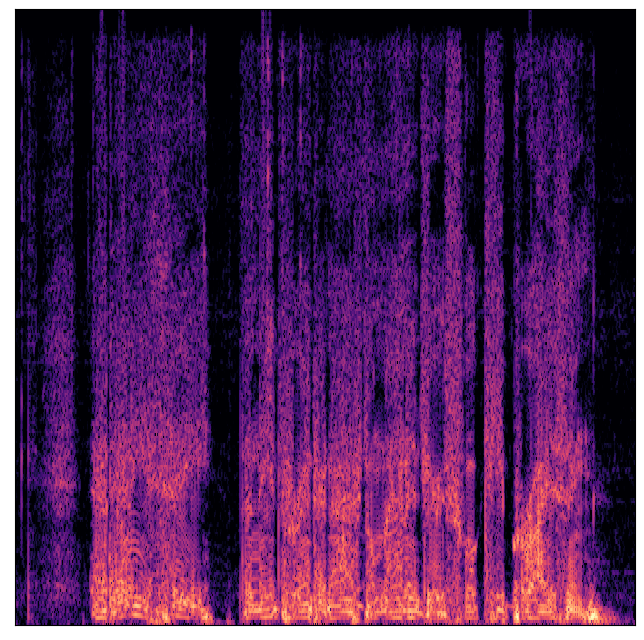};
\end{axis}

\begin{axis}
[
    name={clean2},
    at = {(clean.south west)},
    yshift = \spectroys,
    axis line style={draw=none},
    xmin = 0, xmax = \xmax,
    ymin = 0, ymax = 8000,
    xtick = {0, 1, 2, 3, 4},
    xticklabel style = {xshift=5pt},
    yticklabels = {0, 0, 2, 4, 6},
    yticklabel style = {yshift=5pt},
    xlabel = {Time [s]},
    ylabel = {Frequency [Hz]},
    width =\spectrow,
    height =\spectrow
]
\addplot graphics[xmin=0,ymin=0,xmax=\xmax,ymax=8000] {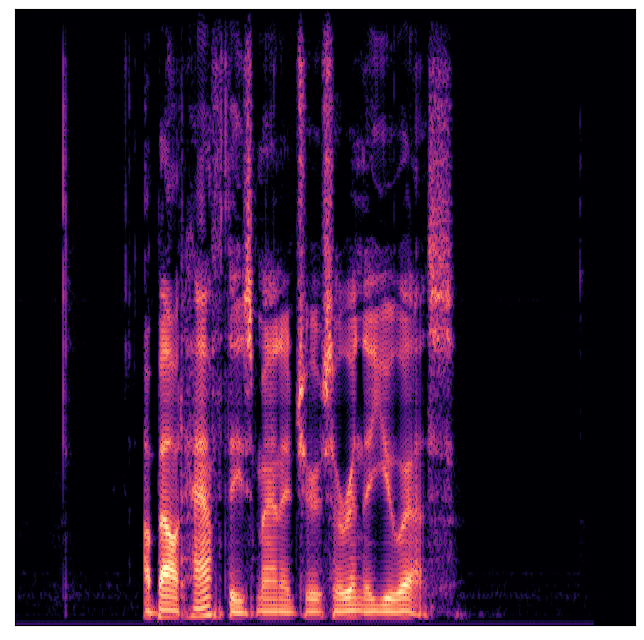};
\end{axis}

\begin{axis}
[
    name={noisy2},
    title = {},
    at = {(clean2.south east)},
    xshift = \spectroxs,
    axis line style={draw=none},
    xmin = 0, xmax = \xmax,
    ymin = 0, ymax = 8000,
    xtick = {0, 1, 2, 3, 4},
    xticklabel style = {xshift=5pt},
    yticklabels=\empty,
    xlabel = {Time [s]},
    width =\spectrow,
    height =\spectrow
]
\addplot graphics[xmin=0,ymin=0,xmax=\xmax,ymax=8000] {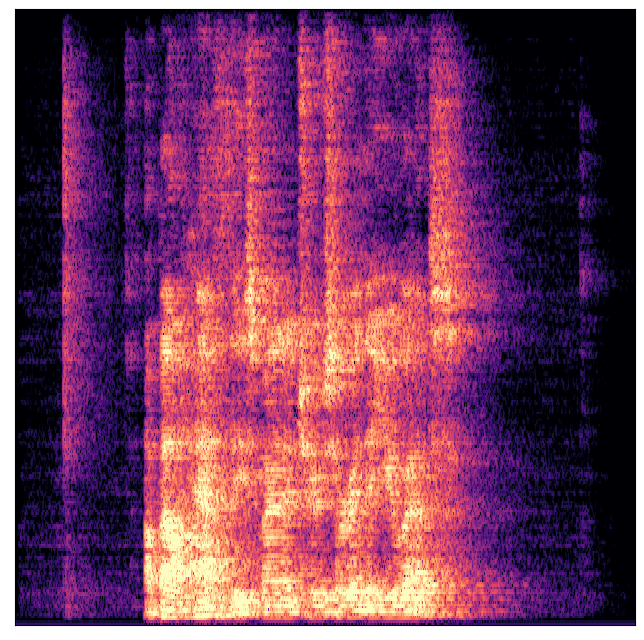};
\end{axis}
 
\begin{axis}
[
    name={predictor2},
    title = {},
    at = {(noisy2.south east)},
    xshift = \spectroxs,
    axis line style={draw=none},
    xmin = 0, xmax = \xmax,
    ymin = 0, ymax = 8000,
    xtick = {0, 1, 2, 3, 4},
    xticklabel style = {xshift=5pt},
    yticklabels=\empty,
    xlabel = {Time [s]},
    width =\spectrow,
    height =\spectrow
]
\addplot graphics[xmin=0,ymin=0,xmax=\xmax,ymax=8000] {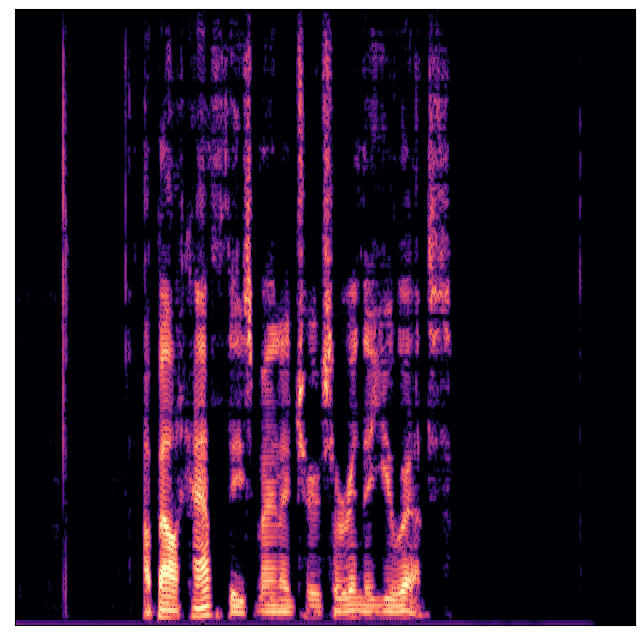};
\end{axis}
 
\begin{axis}
[
    name={residual2},
    title = {},
    at = {(predictor2.south east)},
    xshift = \spectroxs,
    axis line style={draw=none},
    xmin = 0, xmax = \xmax,
    ymin = 0, ymax = 8000,
    xtick = {0, 1, 2, 3, 4},
    xticklabel style = {xshift=5pt},
    yticklabels=\empty,
    xlabel = {Time [s]},
    width =\spectrow,
    height =\spectrow
]
\addplot graphics[xmin=0,ymin=0,xmax=\xmax,ymax=8000] {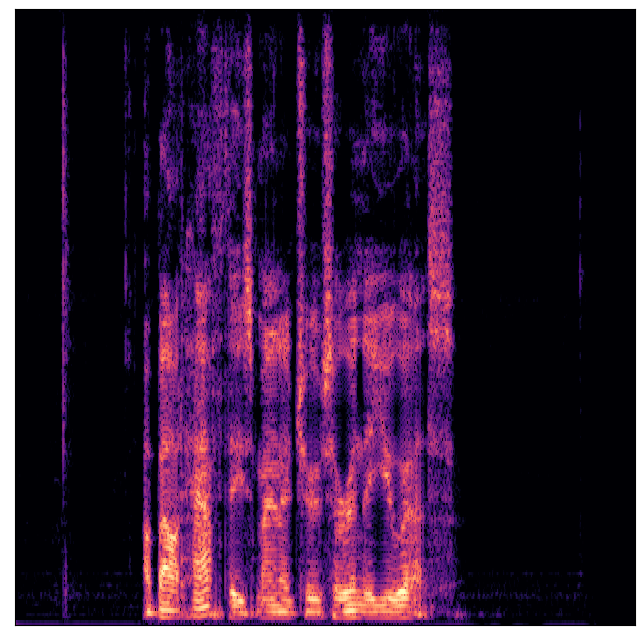};
\end{axis}

\end{tikzpicture}
\vspace{-1em}
\label{fig:residuals}
\end{figure*}

%% file: plots/tikz_stochastic_regeneration_inference.tex
\begin{figure*}

\caption{\centering \textit{Proposed stochastic regeneration inference process. The predictive network is first used to generate a denoised version $D_\theta(\mathbf{y})$. Diffusion-based generation $G_\phi$ is then performed by adding Gaussian noise $\sigma(T)\mathbf{z}$ to obtain the start sample $\mathbf{x}_T$ and solving the reverse diffusion \ac{sde} \eqref{eq:plug-in-reverse-sde}, yielding a sample from the estimated posterior $\mathbf{x}_0 \sim p(\mathbf{x}|D_\theta(\mathbf{y}))$.}}
\centering
\includegraphics[width=0.99\textwidth]{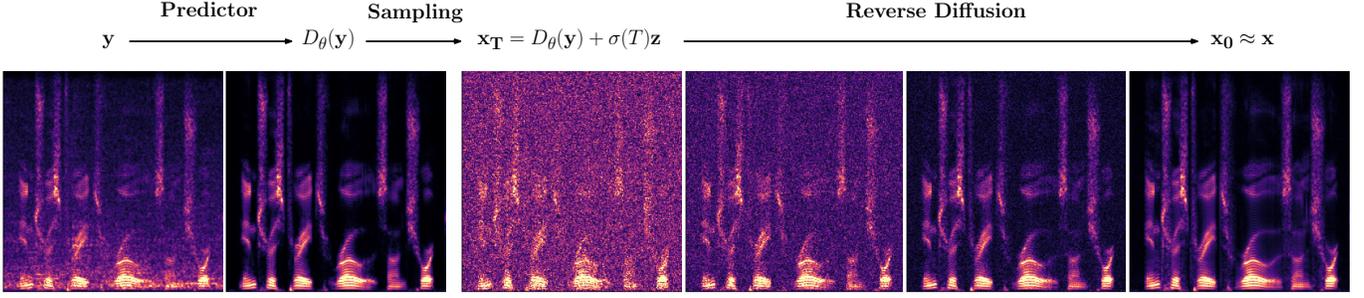}
\label{fig:stochastic-regeneration-inference}

\vspace{-1em}
\end{figure*}

%% file: plots/algo_training.tex
\begin{algorithm}
\caption{StoRM Training}\label{alg:training}
\hspace*{\algorithmicindent} \textbf{Input}: Training set of pairs ($\mathbf{x}, \mathbf{y}$) \\
 \hspace*{\algorithmicindent} \textbf{Output}: Trained parameters $\{ \phi, \theta \}$
\begin{algorithmic}[1]
\setstretch{1.1}
\State Sample diffusion time $t \sim \mathcal{U}(t_\epsilon, T)$
\State Sample noise signal $\mathbf z \sim \mathcal{N}(0, \mathbf{I})$
\State Infer initial prediction $D_\theta(\mathbf{y})$
\State Generate perturbed state $\mathbf x_\tau \leftarrow  \boldsymbol\mu(\mathbf x, D_\theta(\mathbf{y}), \tau) + \sigma(\tau) \mathbf z$
\State Estimate score $\dnnsco (\mathbf{x}_\tau, \left[ \mathbf{y}, D_\theta(\mathbf{y}) \right], \tau)$
\State Compute loss $\mathcal{J}^{(\mathrm{StoRM})}(\phi, \theta)$ (eq.~\eqref{eq:regen-loss})
\State Backpropagate loss $\mathcal{J}^{(\mathrm{StoRM})}(\phi, \theta)$ to update $\{ \phi, \theta \}$

\end{algorithmic}
\end{algorithm}

%% file: plots/algo_inference.tex
\newcommand{\antiv}{\vspace{-0.5em}}

\begin{algorithm}
\caption{StoRM Inference (based on PC sampling by \cite{song2021sde})}\label{alg:inference}
\hspace*{\algorithmicindent} \textbf{Input}: Corrupted speech $\mathbf{y}$, 
step size $\Delta \tau = \frac{T}{N}$ \\
 \hspace*{\algorithmicindent} \textbf{Output}: Clean speech estimate $\widehat{\mathbf{x}}$
\begin{algorithmic}[1]
\setstretch{1.1}
\State Infer initial prediction $D_\theta(\mathbf{y})$
\State Sample noise signal $\mathbf z \sim \mathcal{N}(0, \mathbf{I})$
\State Generate initial reverse state $\mathbf x_T = D_\theta(\mathbf{y}) + \sigma(T) \mathbf z$

\For{$n \in \{ N, \dots, 1 \} $}

\State Get diffusion time $\tau = n \Delta \tau = \frac{n}{N} T$

\If{using corrector}
\State Estimate score $\dnnsco (\mathbf{x}_\tau, \left[ \mathbf{y}, D_\theta(\mathbf{y}) \right], \tau)$
\State Sample correction noise signal $\mathbf w_c \sim \mathcal{N}(0, \mathbf{I})$
\State Correct estimate (Annealed Langevin Dynamics):
\antiv
\begin{align*}
    \mathbf{x}_\tau &\leftarrow \mathbf{x}_\tau + 2 r^2 \sigma(\tau)^2 \dnnsco(\mathbf{x}_\tau, \left[ \mathbf{y}, D_\theta(\mathbf{y}) \right], \tau) \\ &+ 2 r \sigma(\tau) \mathbf{w}_c 
\end{align*}
\EndIf
\antiv

\State Estimate score $\dnnsco(\mathbf{x}_\tau, \left[ \mathbf{y}, D_\theta(\mathbf{y}) \right], \tau)$
\State Sample prediction noise signal $\mathbf w_p \sim \mathcal{N}(0, \mathbf{I})$
\State Predict next Euler-Maruyama step:
\antiv
\begin{align*}\mathbf{x}_{\tau - \Delta\tau} &\leftarrow \mathbf{x}_\tau - \dnnsco (\mathbf{x}_\tau, \left[ \mathbf{y}, D_\theta(\mathbf{y}) \right], \tau) \Delta \tau \\ &+ \gamma (D_\theta(\mathbf{y}) - \mathbf{x}_\tau ) \Delta \tau +  g(\tau) \mathbf{w}_p \sqrt{\Delta \tau}
\end{align*}
\antiv
\EndFor
\antiv

\State Output estimate: $\widehat{\mathbf{x}} = \mathbf{x}_0$

\end{algorithmic}
\end{algorithm}

%% file: plots/clouds.tex
\definecolor{uhhblue}{RGB}{0,156,209}
\definecolor{uhhgreen}{RGB}{66, 178, 60}
\definecolor{uhhred}{RGB}{226,0,26}
\definecolor{uhhblack}{RGB}{0,0,0}
\definecolor{uhhstone}{RGB}{59,81,91}

\begin{figure*}
    \caption{\protect\centering\textit{Visualization of the inference process for the predictive, generative and proposed StoRM models for a complex posterior distribution (see also Figure~1 in \cite{delbracio2023inversion}. With the proposed two-stage inference, StoRM uses the predictive mapping to the posterior mean $\mathbb{E}[\mathbf x | \mathbf y]$ as an intermediate step for easier generative inference of a posterior sample $\mathbf x$ which is more likely to lie in high-density regions of the posterior $p(\mathbf x | \mathbf y)$}
    \vspace{-0.5em}
    }
    \centering
    \begin{minipage}{.32\textwidth}
    \centering
    \input{plots/clouds_predictive}
    Predictive approach
    \end{minipage}
    \begin{minipage}{.32\textwidth}
    \centering
    \input{plots/clouds_generative}
    Generative model
    \end{minipage}
    \begin{minipage}{.32\textwidth}
    \centering
    \input{plots/clouds_storm}
    StoRM
    \end{minipage}
    \label{fig:clouds}
    \vspace{-1em}
\end{figure*}
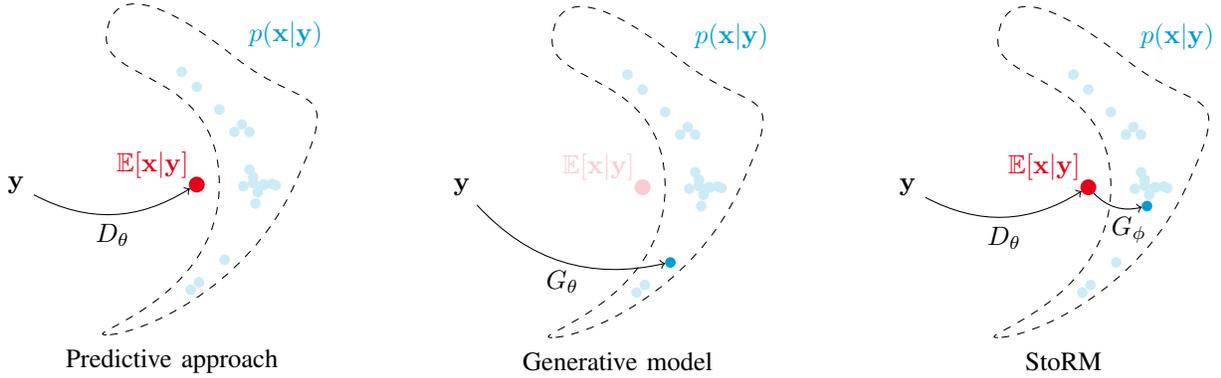

%% file: plots/clouds_predictive.tex
\tikzset{
dot/.style = {circle, fill, minimum size=#1,
              inner sep=0pt, outer sep=0pt}
}

\begin{tikzpicture}
    \draw[smooth cycle, tension=1.4, style=dashed] plot coordinates{(2.5,0) (0,-2) (1.5, 0) (0,2) (1.7, 1.7) } node  [below,label={[label distance=0pt]60:\textcolor{uhhblue}{$p(\mathbf{x}|\mathbf{y})$}}] {};

    \node at (-1.2, 0) (y) {$\mathbf{y}$};

    \node [dot=6pt, fill=uhhred, label={[label distance=-4pt]120:\textcolor{uhhred}{$\mathbb{E}[\mathbf{x}|\mathbf{y}]$}}] at (1.2, 0) (mean) {};

    \node [dot=4pt, fill=uhhblue!20] at (1.0,1.5) (point1) {};
    
    \node [dot=4pt, fill=uhhblue!20] at (1.2, 1.3) (point2) {};
    \node [dot=4pt, fill=uhhblue!20] at (1.5, 1.0) (point3) {};
    \node [dot=4pt, fill=uhhblue!20] at (1.8, 0.8) (point4) {};
    \node [dot=4pt, fill=uhhblue!20] at (1.9, 0.7) (point4) {};
    \node [dot=4pt, fill=uhhblue!20] at (1.7, 0.7) (point4) {};
    \node [dot=4pt, fill=uhhblue!20] at (1.9, 0.2) (point4) {};
    \node [dot=4pt, fill=uhhblue!20] at (1.95, 0.1) (point5) {};
    \node [dot=4pt, fill=uhhblue!20] at (1.95, 0.02) (point5) {};
    \node [dot=4pt, fill=uhhblue!20] at (2.1, 0.0) (point5) {};
    \node [dot=4pt, fill=uhhblue!20] at (2.2, 0.01) (point5) {};

    \node [dot=4pt, fill=uhhblue!20] at (1.12,-1.4) (point2) {};
    \node [dot=4pt, fill=uhhblue!20] at (1.23, -1.3) (point2) {};
    \node [dot=4pt, fill=uhhblue!20] at (1.57, -1.0) (point3) {};
    \node [dot=4pt, fill=uhhblue!20] at (1.98, -0.25) (pointX) {};
    \node [dot=4pt, fill=uhhblue!20] at (2.01, -0.1) (point5) {};
    \node [dot=4pt, fill=uhhblue!20] at (1.82, -0.02) (point5) {};
    \node [dot=4pt, fill=uhhblue!20] at (2.08, -0.0) (point5) {};
    \node [dot=4pt, fill=uhhblue!20] at (2.25, -0.01) (point5) {};

    \path[->] (y) edge [bend right] node [below] {$D_\theta$} (mean);
\end{tikzpicture}

%% file: plots/clouds_generative.tex
\tikzset{
dot/.style = {circle, fill, minimum size=#1,
              inner sep=0pt, outer sep=0pt}
}

\begin{tikzpicture}
    \draw[smooth cycle, tension=1.4, style=dashed] plot coordinates{(2.5,0) (0,-2) (1.5, 0) (0,2) (1.7, 1.7) } node  [below,label={[label distance=0pt]60:\textcolor{uhhblue}{$p(\mathbf{x}|\mathbf{y})$}}] {};

    \node at (-1.2, 0) (y) {$\mathbf{y}$};

    \node [dot=6pt, fill=uhhred!20, label={[label distance=-4pt]120:\textcolor{uhhred!20}{$\mathbb{E}[\mathbf{x}|\mathbf{y}]$}}] at (1.2, 0) (mean) {};

    \node [dot=4pt, fill=uhhblue!20] at (1.0,1.5) (point1) {};
    
    \node [dot=4pt, fill=uhhblue!20] at (1.2, 1.3) (point2) {};
    \node [dot=4pt, fill=uhhblue!20] at (1.5, 1.0) (point3) {};
    \node [dot=4pt, fill=uhhblue!20] at (1.8, 0.8) (point4) {};
    \node [dot=4pt, fill=uhhblue!20] at (1.9, 0.7) (point4) {};
    \node [dot=4pt, fill=uhhblue!20] at (1.7, 0.7) (point4) {};
    \node [dot=4pt, fill=uhhblue!20] at (1.9, 0.2) (point4) {};
    \node [dot=4pt, fill=uhhblue!20] at (1.95, 0.1) (point5) {};
    \node [dot=4pt, fill=uhhblue!20] at (1.95, 0.02) (point5) {};
    \node [dot=4pt, fill=uhhblue!20] at (2.1, 0.0) (point5) {};
    \node [dot=4pt, fill=uhhblue!20] at (2.2, 0.01) (point5) {};

    \node [dot=4pt, fill=uhhblue!20] at (1.12,-1.4) (point2) {};
    \node [dot=4pt, fill=uhhblue!20] at (1.23, -1.3) (point2) {};
    \node [dot=4pt, fill=uhhblue] at (1.57, -1.0) (pointX) {};
    \node [dot=4pt, fill=uhhblue!20] at (1.98, -0.25) (point4) {};
    \node [dot=4pt, fill=uhhblue!20] at (2.01, -0.1) (point5) {};
    \node [dot=4pt, fill=uhhblue!20] at (1.82, -0.02) (point5) {};
    \node [dot=4pt, fill=uhhblue!20] at (2.08, -0.0) (point5) {};
    \node [dot=4pt, fill=uhhblue!20] at (2.25, -0.01) (point5) {};

    \path[->] (y) edge [bend right] node [below] {$G_\theta$} (pointX);
    
\end{tikzpicture}

%% file: plots/clouds_storm.tex
\tikzset{
dot/.style = {circle, fill, minimum size=#1,
              inner sep=0pt, outer sep=0pt}
}

\begin{tikzpicture}
    \draw[smooth cycle, tension=1.4, style=dashed] plot coordinates{(2.5,0) (0,-2) (1.5, 0) (0,2) (1.7, 1.7) } node  [below,label={[label distance=0pt]60:\textcolor{uhhblue}{$p(\mathbf{x}|\mathbf{y})$}}] {};

    \node at (-1.2, 0) (y) {$\mathbf{y}$};

    \node [dot=6pt, fill=uhhred, label={[label distance=-4pt]120:\textcolor{uhhred}{$\mathbb{E}[\mathbf{x}|\mathbf{y}]$}}] at (1.2, 0) (mean) {};

    \node [dot=4pt, fill=uhhblue!20] at (1.0,1.5) (point1) {};
    
    \node [dot=4pt, fill=uhhblue!20] at (1.2, 1.3) (point2) {};
    \node [dot=4pt, fill=uhhblue!20] at (1.5, 1.0) (point3) {};
    \node [dot=4pt, fill=uhhblue!20] at (1.8, 0.8) (point4) {};
    \node [dot=4pt, fill=uhhblue!20] at (1.9, 0.7) (point4) {};
    \node [dot=4pt, fill=uhhblue!20] at (1.7, 0.7) (point4) {};
    \node [dot=4pt, fill=uhhblue!20] at (1.9, 0.2) (point4) {};
    \node [dot=4pt, fill=uhhblue!20] at (1.95, 0.1) (point5) {};
    \node [dot=4pt, fill=uhhblue!20] at (1.95, 0.02) (point5) {};
    \node [dot=4pt, fill=uhhblue!20] at (2.1, 0.0) (point5) {};
    \node [dot=4pt, fill=uhhblue!20] at (2.2, 0.01) (point5) {};

    \node [dot=4pt, fill=uhhblue!20] at (1.12,-1.4) (point2) {};
    \node [dot=4pt, fill=uhhblue!20] at (1.23, -1.3) (point2) {};
    \node [dot=4pt, fill=uhhblue!20] at (1.57, -1.0) (point3) {};
    \node [dot=4pt, fill=uhhblue] at (1.98, -0.25) (pointX) {};
    \node [dot=4pt, fill=uhhblue!20] at (2.01, -0.1) (point5) {};
    \node [dot=4pt, fill=uhhblue!20] at (1.82, -0.02) (point5) {};
    \node [dot=4pt, fill=uhhblue!20] at (2.08, -0.0) (point5) {};
    \node [dot=4pt, fill=uhhblue!20] at (2.25, -0.01) (point5) {};

    \path[->] (y) edge [bend right] node [below] {$D_\theta$} (mean);
    \path[->] (mean) edge [bend right] node [below,xshift=0.5em] {$G_\phi$} (pointX);
\end{tikzpicture}

%% file: sections/exp.tex
\subsection{Data}

\hspace{0.25cm} \textit{a) Speech Enhancement:} \vspace{0.35em}
    
    The WSJ0+Chime dataset is generated using clean speech from the WSJ0 corpus \cite{datasetWSJ0} and noise signals from the CHiME3 dataset \cite{barker2015third}. The mixture signal is created by randomly selecting a noise file and adding it to a clean utterance with a \ac{snr} sampled uniformly between -6 and 14$\,$dB.

    The TIMIT+Chime dataset is similarly generated as WSJ0+Chime, using TIMIT as the clean speech corpus \cite{Garofolo1992TIMIT}. We use this dataset for \ac{asr} as oracle annotations are available for \ac{wer} evaluation.

    The VoiceBank/DEMAND dataset is a classical benchmark dataset for speech enhancement using clean speech from the VCTK corpus \cite{valentini2016investigating} excluding two speakers. The utterances are corrupted by recorded noise from the DEMAND database \cite{Thiemann2013DEMAND} and two artificial noise types (babble and speech shaped) at SNRs of 0, 5, 10, and 15 dB for training and validation. The SNR levels of the test set are 2.5, 7.5, 12.5, and 17.5~dB. \vspace{0.5em}

\hspace{0.25cm} \textit{b) Speech Dereverberation:} The WSJ0+Reverb dataset is generated using clean speech data from the WSJ0 dataset and convolving each utterance with a simulated \ac{rir}. We use the \texttt{\small pyroomacoustics}
engine \cite{Scheibler2018PyRoom} to simulate the \acp{rir}. The reverberant room is modeled by sampling uniformly a target $T_\mathrm{60}$ between 0.4 and 1.0 seconds and a room length, width and height in [5,15]$\times$[5,15]$\times$[2,6]~m.  This results in an average \ac{drr} of around -9$\,$dB and \emph{measured} $T_\mathrm{60}$ of 0.91$\,$s. A dry version of the room is generated using the same geometric parameters with a fixed absorption coefficient of 0.99, to generate the corresponding anechoic target.
\vspace{-1em}

\subsection{Hyperparameters and training configuration}
\label{sec:exp:hyperparameters}

\paragraph{Data representation}
Utterances are transformed using a \ac{stft} with a window size of 510, a hop length of 128 and a square-root Hann window, at a sampling rate of $16$kHz. A square-root magnitude warping is used to compress the dynamical range of the input spectrograms \cite{Richter2022SGMSE++}.
For training, sequences of 256 \ac{stft} frames ($\approx$2s) are randomly extracted from the full-length utterances and normalized by the maximum absolute value of the noisy utterance before being fed to the network. 
\vspace{0.5em}

\paragraph{Forward and reverse diffusion}
For all diffusion models, similar values are chosen to parameterize the forward and reverse stochastic processes. The stiffness parameter is fixed to $\gamma=$1.5, the extremal noise levels to $\sigma_\mathrm{min}=$~0.05 and $\sigma_\mathrm{max}=$~0.5, and the extremal diffusion times to $T=1$ and $\tau_\epsilon=$~0.03 as in \cite{Richter2022SGMSE++}. 
Unless stated otherwise (that is, for all results except those in Figure~\ref{fig:results:enh:steps}), $N=50$ time steps are used for reverse diffusion and we adopt the predictor-corrector scheme \cite{song2021sde} with one step of annealed Langevin dynamics correction and a step size of $r=$~0.5.
\vspace{0.5em}

\paragraph{Network architecture} \label{sec:network}
The backbone architecture we use is a lighter configuration of the NCSN++ architecture variant proposed in \cite{song2021sde}, which was used in our previous study \cite{Lemercier2022icassp} and denoted as \textit{NCSN++M}. The following modifications are carried on the up/down-sampling paths of the network: the attention layers are removed (we keep attention in the bottleneck), the number of layers in each encoder-decoder path is decreased from 7 to 4, and only one ResNet block is used per layer instead of two.
This results in a network capacity of roughly 27.8M parameters instead of 65M, without significant degradation of the speech enhancement performance, be it for predictive or generative modelling.

When this NCSN++M configuration is used for score estimation in SGMSE+ \cite{Richter2022SGMSE++}, we call the resulting approach \textit{SGMSE+M}. There, the noisy speech spectrogram $\mathbf y$ and the current diffusion process estimate $\mathbf x_\tau$ real and imaginary channels are stacked and fed to the network as input, and the current noise level $\sigma(\tau)$ is provided as a conditioner. For our proposed approach StoRM, the initial prediction $D_\theta(\mathbf{y})$ is also stacked together with $\mathbf{y}$ and $\mathbf{x_\tau}$: the influence of this double conditioning is examined in an ablation study in Section~\ref{sec:results:ablations}.
For the predictive approach, denoted directly as \textit{NCSN++M}, the noise-conditioning layers are removed and only the noisy speech spectrogram real and imaginary channels are used. This ablation removes only 1.8\% of the original number of parameters, which hardly modifies the network capacity.

We also use ConvTasNet \cite{Luo2018} and GaGNet \cite{Li2022GaGNet} (see next subsection) as alternative initial predictors for StoRM. We train using NCSN++M as the initial predictor and swap it during inference with one of the two networks mentioned above, in order to test the robustness of our proposed stochastic regeneration approach towards unseen predictors (see Section~\ref{sec:results:predictors}.
\vspace{0.5em}

\paragraph{Baselines}
For comparison on WSJ0-based datasets, we compare 
StoRM to the purely generative SGMSE+M and purely predictive NCSN++M. We also report results using the non-causal version of GaGNet \cite{Li2022GaGNet}, a 
predictive denoiser using parallel magnitude- and complex-domain processing in the \ac{tf} domain.
We complement the benchmark on the Voicebank/DEMAND dataset with the non-causal predictive ConvTasNet \cite{Luo2018}, MetricGAN+ \cite{Fu2022MetricGAN+}, MANNER \cite{Park2022Manner}, VoiceFixer \cite{Liu2021VoiceFixer} as well as the generative unsupervised dynamical VAE (DVAE) \cite{Bie2022DVAE}, conditional time-domain diffusion model CDiffuse \cite{lu2022conditional}, stochastic refinement time-domain enhancement scheme SRTNet \cite{Qiu2023SRTNet} and original SGMSE \cite{Welker2022SGMSE}. For all these, we use publicly available code provided by the authors.
\vspace{0.5em}

\paragraph{Training configuration}
We use the Adam optimizer \cite{kingma2015adam} with a learning rate of $10^{-4}$ and an effective batch size of 16. We track an exponential moving average of the DNN weights with a decay of 0.999 to be used for sampling, as it showed to be very effective \cite{Song2020Improved}. We train \acp{dnn} for a maximum of 1000 epochs using early stopping based on the validation loss with a patience of 10 epochs. All models converged before reaching the maximum number of epochs. The generative approach is trained with the denoising score matching criterion \eqref{eq:training-loss}, and the predictive methods use a simple mean-square error loss on the complex spectrogram. The stochastic regeneration approach uses the combined criterion in \eqref{eq:regen-loss}. The default training strategy is that we pre-train the initial predictor with a simple mean-square error loss, then jointly train the predictor and score networks with \eqref{eq:regen-loss}. Different training strategies are examined in an ablation study in Section~\ref{sec:results:ablations}.

\subsection{Evaluation metrics}
\label{sec:exp:eval}

For instrumental evaluation of the speech enhancement and dereverberation performance with clean test data available, we use intrusive measures such as \ac{pesq} \cite{Rix2001PESQ} to assess speech quality, \ac{estoi} \cite{Jensen2016ESTOI} for intelligibility and \ac{sisdr}, \ac{sisir} and \ac{sisar} 
\cite{Leroux2019SISDR} for noise removal.
As in \cite{Lemercier2022icassp}, we complement our metrics benchmark with WV-MOS \cite{Andreev2022Hifi++}
, which is a DNN-based \ac{mos} estimation, and was used by the authors for reference-free assessment of bandwidth extension or speech enhancement performance.

We also evaluate our proposed approach on \ac{asr}, using NVidia's temporal convolutional network QuartzNet 
\cite{kuchaiev2019nemo} as the speech recognition model, and classical \ac{wer} dynamic programming evaluation with the \texttt{jiwer} Python library\footnote{https://github.com/jitsi/jiwer}. We use the pretrained \texttt{Base-en} 18.9M parameters version of QuartzNet for specialized English speech recognition.

Finally, we organize a medium-scale MUSHRA listening test with 9 participants. We ask the participants to rate 10 samples with a single number representing overall quality, including speech distortion, residual distortions and potential artifacts. We use the \texttt{webMUSHRA}\footnote{https://github.com/audiolabs/webMUSHRA} tool with \texttt{pymushra}\footnote{https://github.com/nils-werner/pymushra} server management. The samples are randomly extracted from the WSJ0+Chime and WSJ0+Reverb test sets, ensuring gender and task balance as well as speaker exclusivity (within a given task, a speaker is used once at most).
The approaches evaluated are the predictive NCSN++M, score-based generative model SGMSE+ and our proposed approach StoRM. The noisy mixture is given as a low anchor, and a supplementary anchor is created by increasing the input SNR by 10$\mathrm{dB}$ in comparison to the noisy mixture.

%% file: sections/results.tex
\subsection{Comparison to baselines}

\input{tables/pure-comparison}

\paragraph{WSJ0+Chime and WSJ0+Reverb}
In tables \ref{tab:results:enh:pure-comparison} and \ref{tab:results:derev:pure-comparison}, we show results of the proposed stochastic regeneration StoRM approach as compared to purely predictive GaGNet and NCSN++M and purely generative SGMSE+M, for denoising on WSJ0+Chime and dereverberation on WSJ0+Reverb. Based on preliminary experiments we confirm that the approach can be successfully trained for joint dereverberation and denoising, and we refer the reader to our web page where audio samples are presented for this joint task. In this work, however, we separate the two tasks to get more insights into the individual performance.

We confirm the results from \cite{Lemercier2022icassp}, which is that predictive NCSN++M and GaGNet provide samples with good interference removal (high SI-SDR) and intelligibility (high ESTOI) but lower quality (lower PESQ and WV-MOS) compared to diffusion-based generative SGMSE+M. This gap is stronger for dereverberation than for denoising as already observed, since the average input \ac{snr} for dereverberation is much lower than for denoising. Also, the reverberation interference being a filtered version of the target speech, the predictive method cannot suppress reverberation without introducing significant distortion, which is particularly audible in NCSN++M and GaGNet results.
The generative SGMSE+M, however, is able to extract the speech cues and directly reconstructs with hardly any reverberation left.

It is generally observed that point-wise measures like SI-SDR, SI-SIR and SI-SAR provide generally worse results for generative models than for predictive models \cite{Whang2021StochasticRefinement, MurphyBook2}. This is because generative models try to estimate the posterior distribution, providing better perceptual metrics, whereas predictive models are implicitly trained to recover the posterior mean and average out the distortions for each point, thus yielding higher point-wise fidelity \cite{Lemercier2022icassp, MurphyBook2}.

We observe that our proposed StoRM associates the best of both the predictive and generative worlds, by producing samples with very high quality like generative SGMSE+M, while being approximately as good with interference removal as the predictive NCSN++M.
Again, the observed gap is more significant for dereverberation, where the proposed StoRM outperforms both SGMSE+M and NCSN++M on all metrics. 
Example spectrograms are displayed on Figure~\ref{fig:spectrograms-enh} and \ref{fig:spectrograms-derev}, for denoising and dereverberation respectively.
\vspace{0.5em}

\paragraph{VoiceBank/DEMAND}

\input{tables/vb-dmd}
We report in Table~\ref{tab:results:vb} results of our StoRM configuration against various state-of-the-art speech enhancement baselines on the VoiceBank/DEMAND benchmark.
The \acp{snr} in Voicebank/DEMAND are always positive and distributed around 10$\mathrm{dB}$, which is not very challenging compared to the conditions in our WSJ0+Chime dataset. Consequently, the gap between SGMSE+ and StoRM on Voicebank/DEMAND is not as large as on WSJ0+Chime, which shows that using the initial predictor is particularly useful in difficult conditions.
In easier environments such as that simulated in Voicebank/DEMAND, diffusion-based generative modelling can take the noisy mixture as the initial condition for reverse diffusion without being further guided.
Still, our proposed method StoRM still slightly outperforms the other generative models on ESTOI, WV-MOS and SI-SDR, setting a new state-of-the-art record for generative models on this benchmark. 

\subsection{Efficient sampling}

\input{plots/lineplot_steps_enh}
We report in Figure~\ref{fig:results:enh:steps} the performance of the SGMSE+M and StoRM schemes as a function of the number of steps used for reverse diffusion.
We additionally provide an estimation of the 
number of \ac{mac} operations per second as measured by the \texttt{python-papi} package.

We observe that StoRM is able to maintain performance at a near-optimal level even using only 10 steps, using the initial predictive estimate as a reasonable guess for further diffusion. In comparison, SGMSE+M performance degrades rapidly as the number of steps decreases. 
Furthermore, StoRM is able to produce very high-quality samples without even needing the Annealed Langevin Dynamics corrector during sampling, whereas SGMSE+M performance dramatically degrades without this corrector. Since each corrector step makes an additional call to the score network, avoiding its use further relaxes the computational complexity. 
StoRM therefore highlights a strong compromise between inference speed and sample quality.
Using StoRM with 20 steps and no corrector produces near-optimal sample quality at a cost of 4.5 $\cdot 10^{11}$ MAC$\cdot \mathrm{s}^{-1}$, versus 2.1 $\cdot 10^{12}$ MAC$\cdot \mathrm{s}^{-1}$ for the optimal SGMSE+M setting (50 steps and Annealed Langevin Dynamics correction). StoRM even outperforms the optimal SGMSE+M setting using 10 steps and no corrector, thus reducing computational complexity by a full order of magnitude.
In our recent work \cite{richter2023challenge}, we proposed to make our diffusion generative model causal to demonstrate its application to real-world scenarios.
\vspace{-1em}

\input{plots/spectros_enh_wnoisy.tex}

\subsection{Generalization to unseen data}

\input{tables/mismatch}

In Table~\ref{tab:results:mismatch_wsj0-0-20}, we examine robustness to mismatched training and test data. The mismatched condition is generated by training on Voicebank/DEMAND and testing on WSJ0+Chime$^\star$. WSJ0+Chime$^\star$ is created in the same fashion as WSJ0+Chime, but the input SNRs range between 0 and 20dB to match the SNR range in Voicebank+DEMAND, and thus constitutes the same benchmark as in \cite{Richter2022SGMSE++}.
We observe that NCSN++M shows a reasonable generalization ability because of its sophisticated network architecture. However, SGMSE+M and StoRM's ability to maintain performance as compared to the matched case is even superior. This shows that StoRM can leverage generative modelling to correct the relative lack of robustness of the first predictive stage to this mismatched condition.

\subsection{Generalization to mismatched predictors} \label{sec:results:predictors}

\input{tables/predictors}
In Table~\ref{tab:results:predictors}, we report results for StoRM using different initial predictors than the one used during training. The approach is trained using the NCSN++M as initial predictor as before, and we test using ConvTasNet \cite{Luo2018} and GaGNet \cite{Li2022GaGNet} as alternative initial predictors by exchanging this predictor during inference.
We observe that the artifacts in GaGNet estimates are of a similar nature than those of NCSN++M, as both approaches process speech in the \ac{tf} domain. StoRM is entirely robust to such a slight mismatch, as indicated by the equivalent performance of using NCSN++M and GaGNet as the initial predictor.
ConvTasNet is a time-domain method using a fully learnt encoder: the speech distortions are then different than those of NCSN++M or GaGNet. Additionally, ConvTasNet's original performance is slightly worse than its two counterparts.
Consequently, we observe that the performance of StoRM using ConvTasNet as the initial predictor is poorer but close to that of using NCSN++M as the predictor. This demonstrates relative robustness to unseen conditions provided by the generative modelling stage.

\subsection{ASR results}

\input{plots/barplot_asr}
In Figure~\ref{fig:results:asr}, we compare the predictive, generative and stochastic regeneration approaches on speech enhancement for \ac{asr} using the TIMIT+Chime dataset. We observe that SGMSE+M results in poorer speech recognition abilities than its predictive GaGNet and NCSN++M counterparts, and hardly improves the ASR performance over the noisy mixture. This can be explained by the previously mentioned undesired vocalizing artifacts and phonetic confusions, which are created by the generative approach under uncertainty over the presence and phonetic nature of speech respectively and are heavily punished by \ac{wer} evaluation. Using the predictive estimate as a guide for generation, StoRM improves the \ac{wer} performance by a relative factor of 19\% as compared to SGMSE+M, even slightly outperforming NCSN++M, which shows that most of the artifacts and confusions are corrected.

\subsection{Listening experiment}

\input{plots/boxplot_mushra}
We show in the boxplot on Figure~\ref{fig:results:mushra} the results of our MUSHRA listening test. On average, the participants clearly rated the proposed StoRM higher than the purely predictive NCSN++M and purely generative SGMSE+M. This confirms the results provided by the intrusive and non-intrusive metrics provided in tables \ref{tab:results:enh:pure-comparison} and \ref{tab:results:derev:pure-comparison}. Participants rated NCSN++M slightly better than SGMSE+M on average, which is linked to the rating criterion described in Section~\ref{sec:exp:eval}. This seems to indicate that participants put more weight on "residual distortions" and "potential artifacts" (vocalizing/breathing/confusions) than on "speech distortion".

\subsection{Ablation studies}
\label{sec:results:ablations}

We conduct ablation studies on the WSJ0+Chime dataset, to observe the respective influence of score network conditioning on the one hand and the training strategy on the other hand.

\paragraph{Conditioning of the score network}
\input{tables/conditioning}
In Table~\ref{tab:results:conditioning}, we report instrumental results when using different conditioning inputs for the score network used in the proposed StoRM. We input either the noisy speech $\mathbf{y}$ ("Noisy"), the denoised estimate $D_\theta(\mathbf{y})$ ("PostDenoiser"), or both ("Both", which is the default setting for StoRM).
Using only the noisy speech ("Noisy") is detrimental to the performance. It seems that the score network does need the information from the original distortions in $D_\theta(\mathbf{y})$ at time step $\tau$=$T$, to properly learn the score at time step $\tau$$<$$T$. This mismatch at the first denoising steps is detrimental to performance. 
We also observe that instrumental metrics tend to slightly favor the "Both" conditioning over the "PostDenoiser" conditioning.

\paragraph{Training strategies}

\input{tables/training.tex}
We show in Table~\ref{tab:results:training} the results of StoRM using different training strategies. We see that jointly training the initial predictor and the score network slightly improves results for denoising. However, training the initial predictor from scratch or having it pre-trained first does not seem to make a difference, as long as one regularizes the training criterion with the supervised criterion $\mathcal{J}^{(\mathrm{Sup})}$ which matches the output of the initial predictor to the target. Indeed, as shown in the third line of Table~\ref{tab:results:training}, if we use a randomly initialized predictor and train both the predictor and score networks only with the score matching criterion $\mathcal{J}^{(\mathrm{DSM})}$--- i.e. setting $\alpha$ in \eqref{eq:regen-loss} to 0---the performance dramatically drops. This is to be expected since the learning task then becomes much more complicated given the size of the search space and the lack of regularization. The proposed combination of joint training and regularization with $\mathcal{J}^{(\mathrm{Sup})}$ performs most favorably. This indicates that it is best to train the predictor to output something resembling clean speech rather than arbitrary learned encoder features, while still leaving some room for the predictor to adapt its output to the score model.
Our experiments also indicate that, once pre-trained, StoRM converges $25\%$ faster than when training from scratch. However, the cumulated time of pre-training the predictor and fine-tuning is $50\%$ larger than the cost of training from scratch.

%% file: tables/pure-comparison.tex
\begin{table*}[t]
    \centering
    \caption{\centering\textit{Denoising results obtained on WSJ0+Chime. Values indicate mean and standard deviation. All approaches (except GaGNet) use the NCSN++M architecture. Diffusion models (SGMSE+M and StoRM) use $N=50$ steps for reverse diffusion.}}
    \scalebox{0.99}{
    \begin{tabular}{c|ccccccc}
    
\toprule 
Method & WV-MOS & 
PESQ & ESTOI & SI-SDR & SI-SIR & SI-SAR \\

\midrule
\midrule
Mixture  & 1.43 $\pm$ 0.66 & 
1.38 $\pm$ 0.32 & 0.65 $\pm$ 0.18 & $\,$$\,$$\,$4.3 $\pm$ 5.8 & $\,$$\,$$\,$4.3 $\pm$ 5.8 & - \\
\midrule

SGMSE+M & 3.63 $\pm$ 0.38 & 
2.33 $\pm$ 0.61 & 0.86 $\pm$ 0.10 & 13.3 $\pm$ 5.0 & 27.4 $\pm$ 6.3 & 13.5 $\pm$ 4.9 \\

NCSN++M & 3.47 $\pm$ 0.53 & 
2.21 $\pm$ 0.65 & \textbf{0.89 $\pm$ 0.09} & \textbf{16.4 $\pm$ 4.4} & 31.1 $\pm$ 5.0 & \textbf{16.6 $\pm$ 4.4} \\

GaGNet & 3.34 $\pm$ 0.54 & 
2.19 $\pm$ 0.61 & 0.87 $\pm$ 0.09 & 15.7 $\pm$ 4.3 & 27.6 $\pm$ 4.7 & 16.0 $\pm$ 4.4 \\
\midrule

StoRM & \textbf{3.72 $\pm$ 0.40} & 
\textbf{2.58 $\pm$ 0.61} & 0.88 $\pm$ 0.08 & 15.1 $\pm$ 4.2 & \textbf{31.6 $\pm$ 5.0} & 15.3 $\pm$ 4.2 \\

 \midrule
    \bottomrule
    \end{tabular}
    }
    \label{tab:results:enh:pure-comparison}
\end{table*}

\begin{table*}[t]
    \centering
    \caption{\centering\textit{Dereverberation results on Reverb-WSJ0. Values indicate mean and standard deviation. All approaches (except GaGNet) use the NCSN++M architecture. Diffusion models (SGMSE+M and StoRM) use $N=50$ steps for reverse diffusion.}}
    \scalebox{0.99}{
    \begin{tabular}{c|cccccc}
    
\toprule 
Method & WV-MOS & 
PESQ & ESTOI & SI-SDR & SI-SIR & SI-SAR \\

\midrule
\midrule
Mixture & 1.78 $\pm$ 0.99 & 
1.36 $\pm$ 0.19 & 0.46 $\pm$ 0.12 & -7.3 $\pm$ 5.5 &  $\,$-7.5 $\pm$ 5.4 & - \\
\midrule

SGMSE+M & 3.49 $\pm$ 0.39 & 
2.66 $\pm$ 0.45 & 0.85 $\pm$ 0.06 & $\,$$\,$2.4 $\pm$ 7.2 & 11.6 $\pm$ 9.9 & $\,$2.8 $\pm$ 6.8 \\

NCSN++M & 2.99 $\pm$ 0.38 & 
2.08 $\pm$ 0.47 & 0.85 $\pm$ 0.06 & $\,$$\,$6.1 $\pm$ 3.8 & 21.4 $\pm$ 7.0 & $\,$6.1 $\pm$ 3.7 \\

GaGNet & 2.40 $\pm$ 0.52 & 1.59 $\pm$ 0.37 & 0.68 $\pm$ 0.09 & -0.5 $\pm$ 4.8 & $\,$$\,$$\,$7.7 $\pm$ 4.0 & $\,$0.2 $\pm$ 5.1 \\
\midrule

StoRM & \textbf{3.73 $\pm$ 0.32} & 
\textbf{2.83 $\pm$ 0.42} & \textbf{0.88 $\pm$ 0.04} & \textbf{6.5 $\pm$ 4.0} & \textbf{22.9 $\pm$ 8.2} & $\,$\textbf{6.5 $\pm$ 3.9} \\

 \midrule
    \bottomrule
    \end{tabular}
    }
    \label{tab:results:derev:pure-comparison}
\end{table*}

%% file: tables/vb-dmd.tex
\begin{table}[t]
    \centering
    \caption{\centering\textit{Denoising results obtained on VoiceBank/DEMAND. 
    $\mathrm{P}$ means predictive and $\mathrm{G}$ generative. All approaches were evaluated on the test set using publicly available code attached to the respective papers.}}
    \hspace{-0.5em}
    \scalebox{0.83}{
    \begin{tabular}{lccccc}
        \toprule 
        Method & Type & PESQ & ESTOI & SI-SDR & WV-MOS \\
        \midrule
        \midrule
        Mixture  & & 1.97  & 0.79 &  $\;\,$8.4 & 2.99\\
        \midrule
        NCSN++ & P & 2.83 & \textbf{0.88} & \textbf{20.1} & 4.07 \\ 
        NCSN++M & P & 2.82 & \textbf{0.88} & 19.9 & 4.06 \\ 
        Conv-TasNet \cite{Luo2018} & P & 2.84 & 0.85 & 19.1 & 4.28 \\
        MetricGAN+ \cite{Fu2022MetricGAN+} & P & 3.13 & 0.83 & $\,$$\,$ 8.5 & 3.90 \\
        MANNER \cite{Park2022Manner} & P & \textbf{3.21} & 0.87 & 18.9 & \textbf{4.38} \\
        GaGNet \cite{Li2022GaGNet} & P & 2.94 & 0.86 & 18.1 & 4.23 \\
        VoiceFixer \cite{Liu2021VoiceFixer} & P & 2.11 & 0.73 & -4.3 & 4.00 \\
        \midrule
        DVAE \cite{Bie2022DVAE} & G & 2.43  & 0.81 & 16.4 & 3.73  \\ 
        CDiffuSE \cite{lu2022conditional} & G & 2.46 & 0.79 & 12.6 & 3.64 \\
        SRTNet \cite{Qiu2023SRTNet} & G & 2.11 & 0.81 & $\,$8.5 & 3.58\\
        SGMSE \cite{Welker2022SGMSE} & G & 2.28  & 0.80 & 16.2 & 3.90 \\
        SGMSE+ \cite{Richter2022SGMSE++} & G & 2.93 & 0.87 & 17.3 & 4.24 \\
        SGMSE+M & G & \textbf{2.96} & 0.87 & 17.3 & 4.26 \\
        StoRM (proposed) & G & 2.93 & \textbf{0.88} & \textbf{18.8} & \textbf{4.30} \\
        \bottomrule
    \end{tabular}
    }
    \vspace{-1em}
    \label{tab:results:vb}
\end{table}

%% file: plots/lineplot_steps_enh.tex
\newcommand{\w}{0.23\textwidth}
\newcommand{\h}{0.23\textwidth}
\newcommand{\xs}{0.05\textwidth}
\newcommand{\ys}{0.02\textwidth}
\newcommand{\lxs}{0.385\textwidth}
\newcommand{\tys}{87pt}
\newcommand{\txs}{30pt}

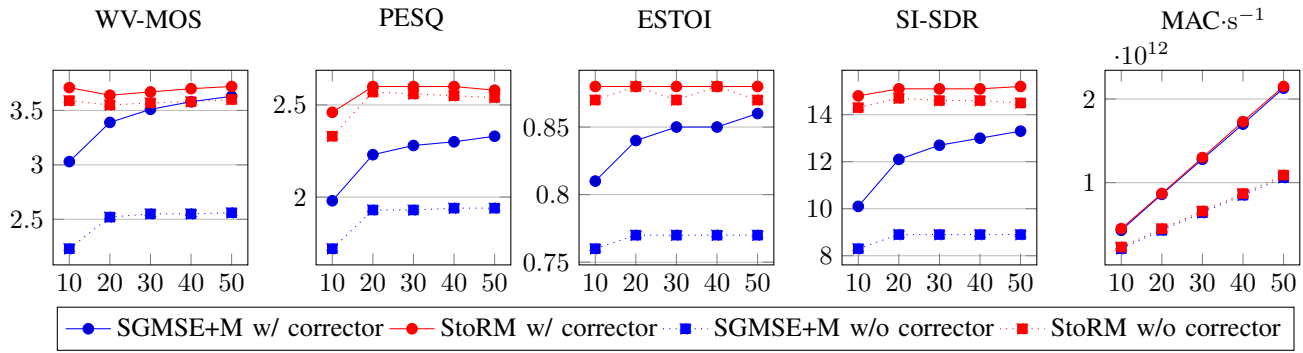
\begin{figure*}
    \caption{\centering \textit{Results for denoising on WSJ0+Chime as a function of the number of reverse diffusion steps $N$. All approaches use the same NCSN++M architecture. The corrector uses one step of Annealed Langevin Dynamics with  $r=0.5$. 
    \vspace{-0.5em}
    }}
    \centering
    
\begin{tikzpicture}

\begin{axis}[title={WV-MOS}, name=wvmos,
ymajorgrids,
width=\w, height=\h, xshift=\xs,
xtick={10,20,..., 50},
legend style={
    at={(xticklabel cs:.5)},
    anchor=north,
    xshift=\lxs,
    yshift=-0.25em,
},
legend columns=5, mark=none,
every axis title/.style={yshift=\tys, xshift=\txs}
]

\addplot+[blue,, mark=*] table [x=N, y=WVMOS, col sep=comma] {sgmse_enh.csv};
\addplot+[red, mark=*, mark options={fill=red},] table [x=N, y=WVMOS, col sep=comma] {regensgmsePJ_enh.csv};

\addplot+[blue,, dotted, mark=square*, mark options={fill=blue}] table [x=N, y=WVMOS, col sep=comma] {sgmse_enh_NOCORR.csv};
\addplot+[red, mark=square*, mark options={fill=red},, dotted] table [x=N, y=WVMOS, col sep=comma] {regensgmsePJ_enh_NOCORR.csv};

\legend{SGMSE+M w/ corrector, StoRM w/ corrector, SGMSE+M w/o corrector, StoRM w/o corrector}
\end{axis}

\begin{axis}[title={PESQ}, name=pesq, at={(wvmos.south east)},
ymajorgrids,
width=\w, height=\h, xshift=\xs,
xtick={10,20,..., 50},
every axis title/.style={yshift=\tys, xshift=\txs}
]

\addplot+[blue,, mark=*] table [x=N, y=PESQ, col sep=comma] {sgmse_enh.csv};
\addplot+[red, mark=*, mark options={fill=red},] table [x=N, y=PESQ, col sep=comma] {regensgmsePJ_enh.csv};

\addplot+[blue,, dotted, mark=square*, mark options={fill=blue}] table [x=N, y=PESQ, col sep=comma] {sgmse_enh_NOCORR.csv};
\addplot+[red, mark=square*, mark options={fill=red},, dotted] table [x=N, y=PESQ, col sep=comma] {regensgmsePJ_enh_NOCORR.csv};

\end{axis}

\begin{axis}[title={ESTOI}, name=estoi, at={(pesq.south east)},
ymajorgrids,
width=\w, height=\h, xshift=\xs,
xtick={10,20,..., 50},
every axis title/.style={yshift=\tys, xshift=\txs}
]

\addplot+[blue, mark=*] table [x=N, y=ESTOI, col sep=comma] {sgmse_enh.csv};
\addplot+[red, mark=*, mark options={fill=red},] table [x=N, y=ESTOI, col sep=comma] {regensgmsePJ_enh.csv};

\addplot+[blue, mark=square*, dotted, mark options={fill=blue}] table [x=N, y=ESTOI, col sep=comma] {sgmse_enh_NOCORR.csv};
\addplot+[red, mark=square*, mark options={fill=red},, dotted] table [x=N, y=ESTOI, col sep=comma] {regensgmsePJ_enh_NOCORR.csv};

\end{axis}

\begin{axis}[title={SI-SDR}, name=sisdr, at={(estoi.south east)},
ymajorgrids,
width=\w, height=\h, xshift=\xs,
xtick={10,20,..., 50},
every axis title/.style={yshift=\tys, xshift=\txs}
]

\addplot+[blue, mark=*] table [x=N, y=SISDR, col sep=comma] {sgmse_enh.csv};
\addplot+[red, mark=*, mark options={fill=red},] table [x=N, y=SISDR, col sep=comma] {regensgmsePJ_enh.csv};

\addplot+[blue, mark=square*, , dotted, mark options={fill=blue}] table [x=N, y=SISDR, col sep=comma] {sgmse_enh_NOCORR.csv};
\addplot[red, mark=square*, mark options={fill=red}, dotted] table [x=N, y=SISDR, col sep=comma] {regensgmsePJ_enh_NOCORR.csv};

\end{axis}

\begin{axis}[title={MAC$\cdot \mathrm{s}^{-1}$}, name=sisdr, at={(sisdr.south east)},
ymajorgrids,
width=\w, height=\h, xshift=\xs,
xtick={10,20,..., 50},
every axis title/.style={yshift=\tys, xshift=35pt}
]

\addplot+[blue, mark=*] table [x=N, y=MMAC, col sep=comma] {sgmse_enh.csv};
\addplot+[red, mark=*, mark options={fill=red},] table [x=N, y=MMAC, col sep=comma] {regensgmsePJ_enh.csv};

\addplot+[blue, mark=square*, , dotted, mark options={fill=blue}] table [x=N, y=MMAC, col sep=comma] {sgmse_enh_NOCORR.csv};
\addplot[red, mark=square*, mark options={fill=red}, dotted] table [x=N, y=MMAC, col sep=comma] {regensgmsePJ_enh_NOCORR.csv};

\end{axis}

\end{tikzpicture}
    \label{fig:results:enh:steps}
\end{figure*}

%% file: plots/spectros_enh_wnoisy.tex
\renewcommand{\spectrow}{.27\textwidth}
\renewcommand{\spectroxs}{.011\textwidth}
\renewcommand{\xmax}{6}

\begin{figure*}

\caption{\protect\centering \textit{Clean, noisy and processed utterances from WS0+Chime. Input \ac{snr} is -0.9~$\mathrm{dB}$. Vocalizing artifacts are visible at the beginning of the SGMSE+M utterance. Speech distortions are observed in the NCSN++M sample. StoRM corrects these distortions without introducing vocalizing artifacts and yield high quality and intelligibility.
 }}
\begin{tikzpicture}[scale=0.95, transform shape]
 \protect\centering
\begin{axis}
[
    name={clean},
    title = {Clean},
    axis line style={draw=none},
    xmin = 0, xmax = \xmax,
    ymin = 0, ymax = 8000,
    xtick = {0, 1, 2, 3, 4, 5},
    xticklabel style = {xshift=5pt},
    ytick = {0, 2000, ..., 8000},
    yticklabel style = {yshift=5pt},
    yticklabels={0, 2, 4, 6},
    xlabel = {Time [s]},
    ylabel = {Frequency [kHz]},
    width =\spectrow,
    height =\spectrow
]
\addplot graphics[xmin=0,ymin=0,xmax=\xmax,ymax=8000] {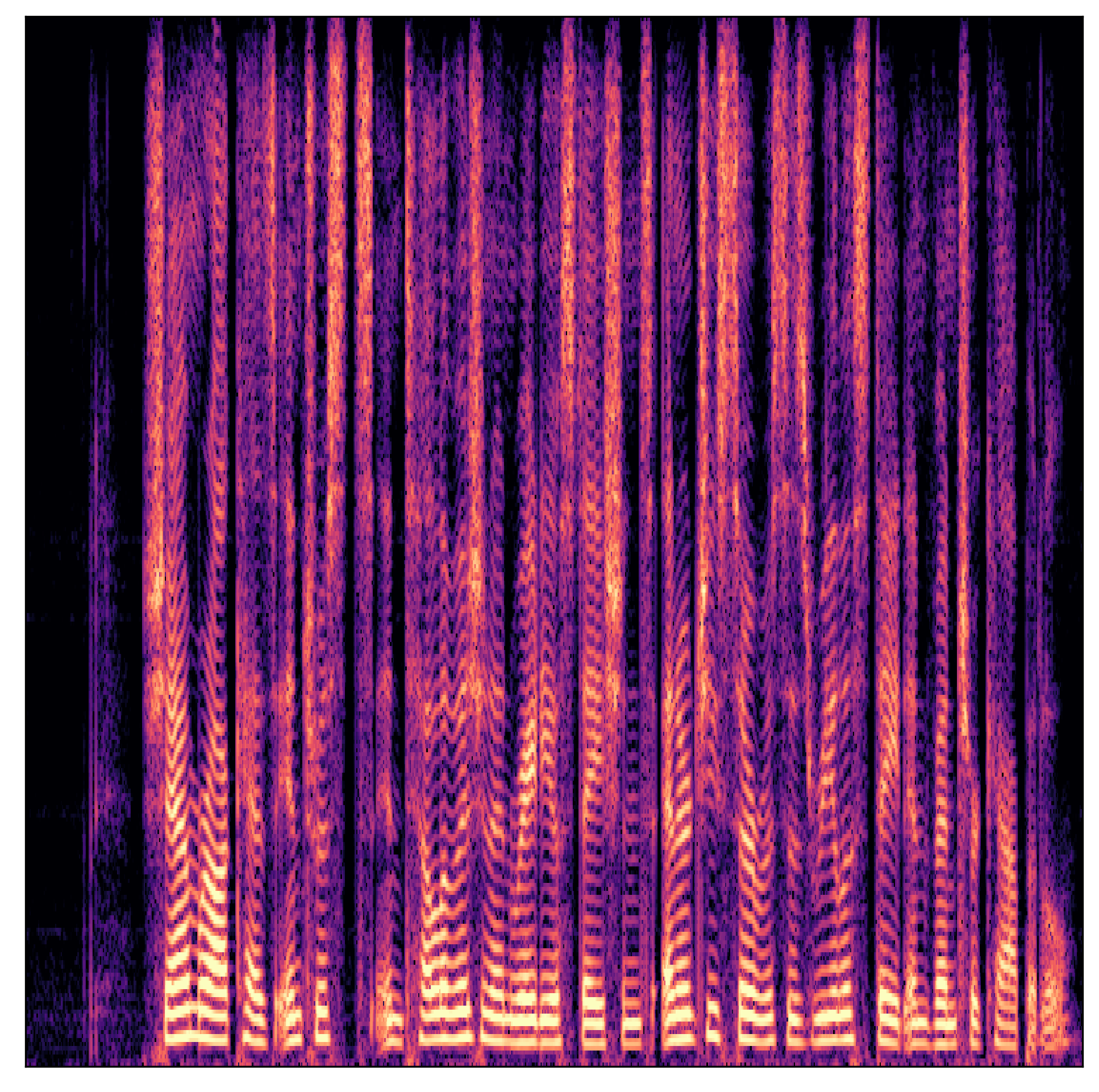};
\end{axis}

\begin{axis}
[
    name = {noisy},
    title = {Noisy},
    at={(clean.south east)},
    xshift = \spectroxs,
    axis line style={draw=none},
    xmin = 0, xmax = \xmax,
    ymin = 0, ymax = 8000,
    xtick = {0, 1, 2, 3, 4, 5},
    xticklabel style = {xshift=5pt},
    yticklabels=\empty,
    xlabel = {Time [s]},
    width =\spectrow,
    height =\spectrow,
]
\addplot graphics[xmin=0,ymin=0,xmax=\xmax,ymax=8000] {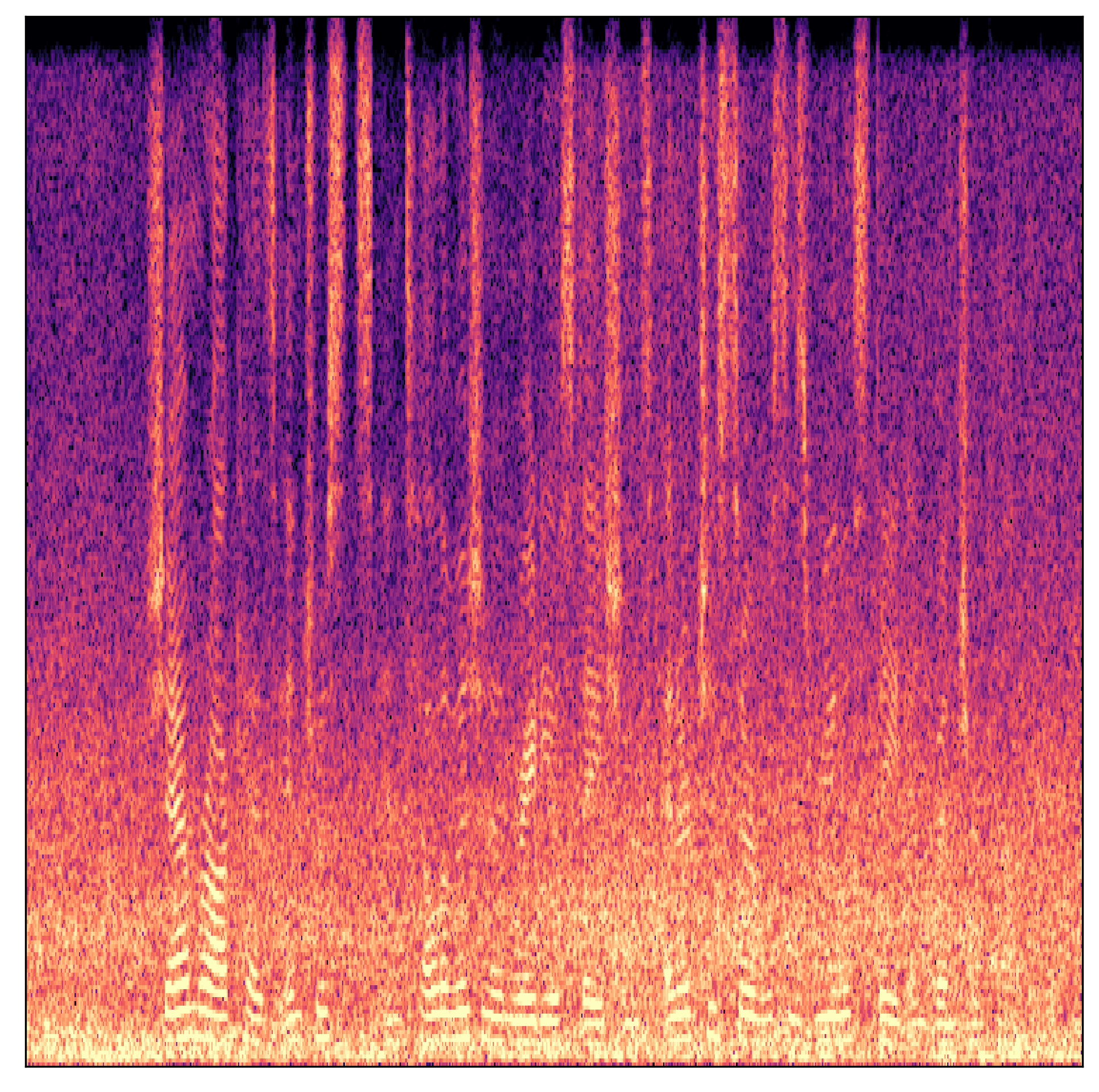};
\end{axis}

\begin{axis}
[
    name = {ncsn},
    title = {NCSN++M},
    at={(noisy.south east)},
    xshift = \spectroxs,
    axis line style={draw=none},
    xmin = 0, xmax = \xmax,
    ymin = 0, ymax = 8000,
    xtick = {0, 1, 2, 3, 4, 5},
    xticklabel style = {xshift=5pt},
    yticklabels=\empty,
    xlabel = {Time [s]},
    width =\spectrow,
    height =\spectrow,
]
\addplot graphics[xmin=0,ymin=0,xmax=\xmax,ymax=8000] {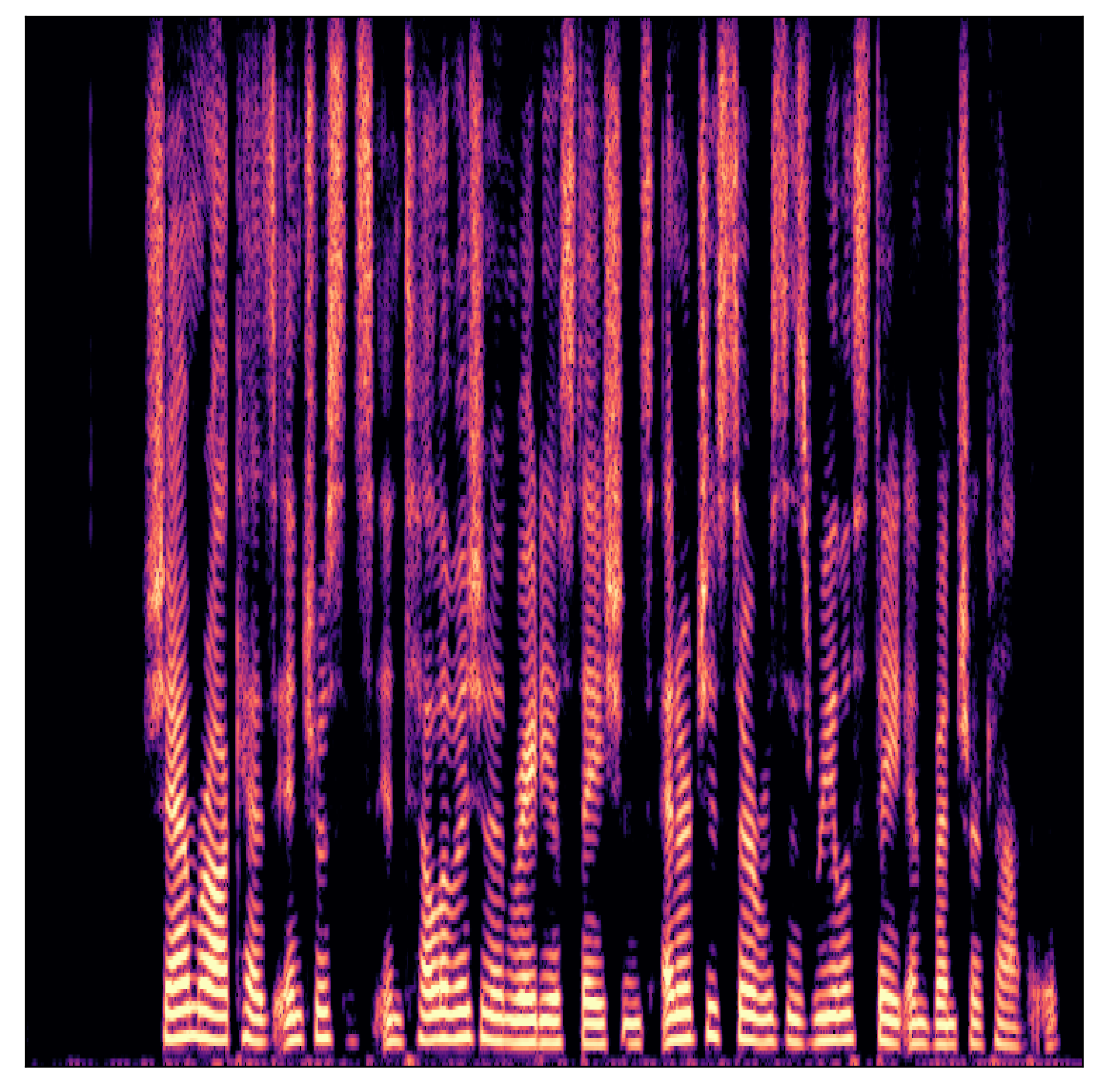};

\end{axis}

\begin{axis}
[
    name = {sgmse},
    title = {SGMSE+M},
    at={(ncsn.south east)},
    xshift = \spectroxs,
    axis line style={draw=none},
    xmin = 0, xmax = \xmax,
    ymin = 0, ymax = 8000,
    xtick = {0, 1, 2, 3, 4, 5},
    xticklabel style = {xshift=5pt},
    yticklabels=\empty,
    xlabel = {Time [s]},
    width =\spectrow,
    height =\spectrow,
]
\addplot graphics[xmin=0,ymin=0,xmax=\xmax,ymax=8000] {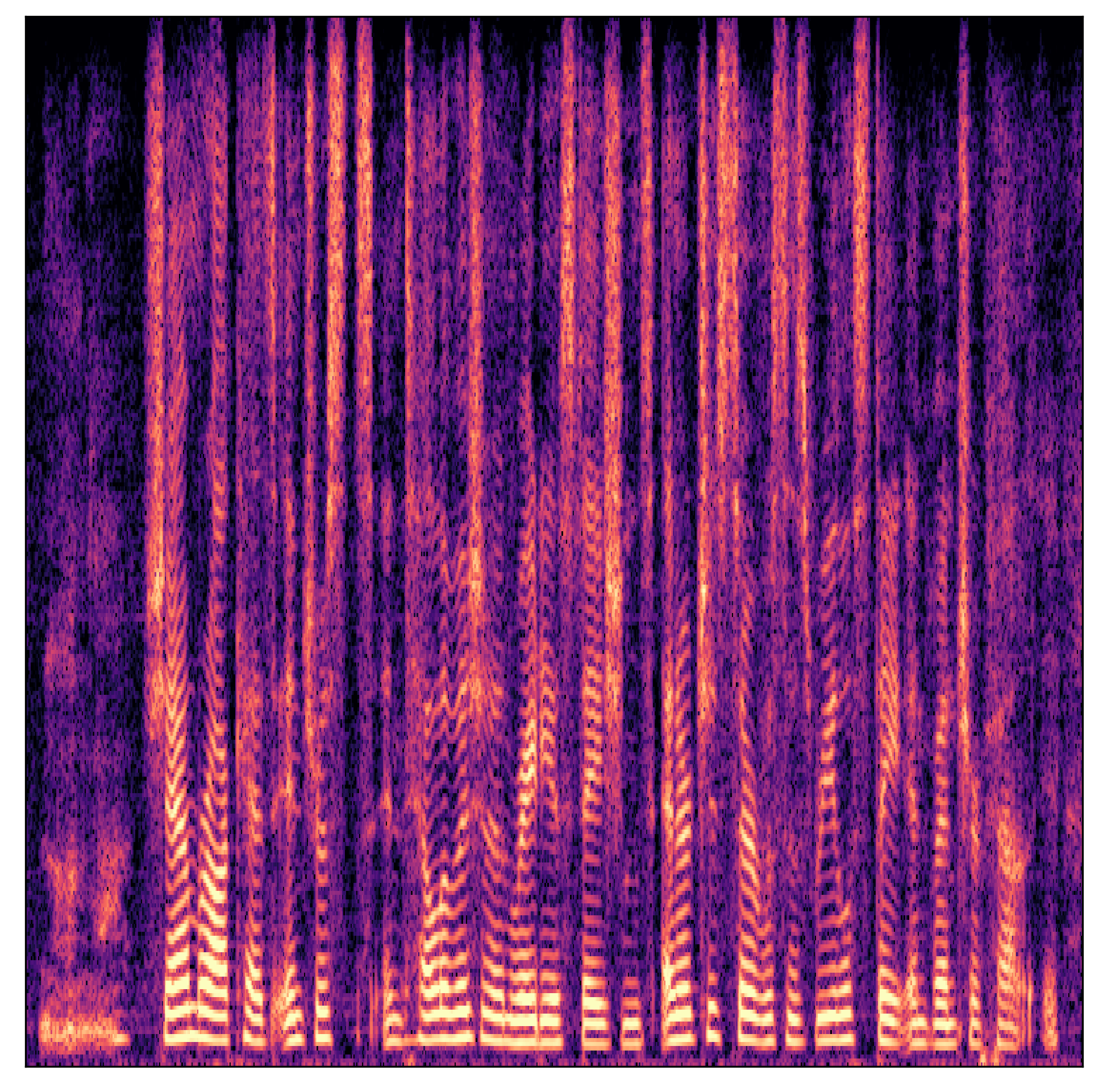};

\end{axis}

\begin{axis}
[
    name = {regensgmse},
    title = {StoRM},
    at={(sgmse.south east)},
    xshift = \spectroxs,
    axis line style={draw=none},
    xmin = 0, xmax = \xmax,
    ymin = 0, ymax = 8000,
    xtick = {0, 1, 2, 3, 4, 5},
    xticklabel style = {xshift=5pt},
    yticklabels=\empty,
    xlabel = {Time [s]},
    width =\spectrow,
    height =\spectrow,
]
\addplot graphics[xmin=0,ymin=0,xmax=\xmax,ymax=8000] {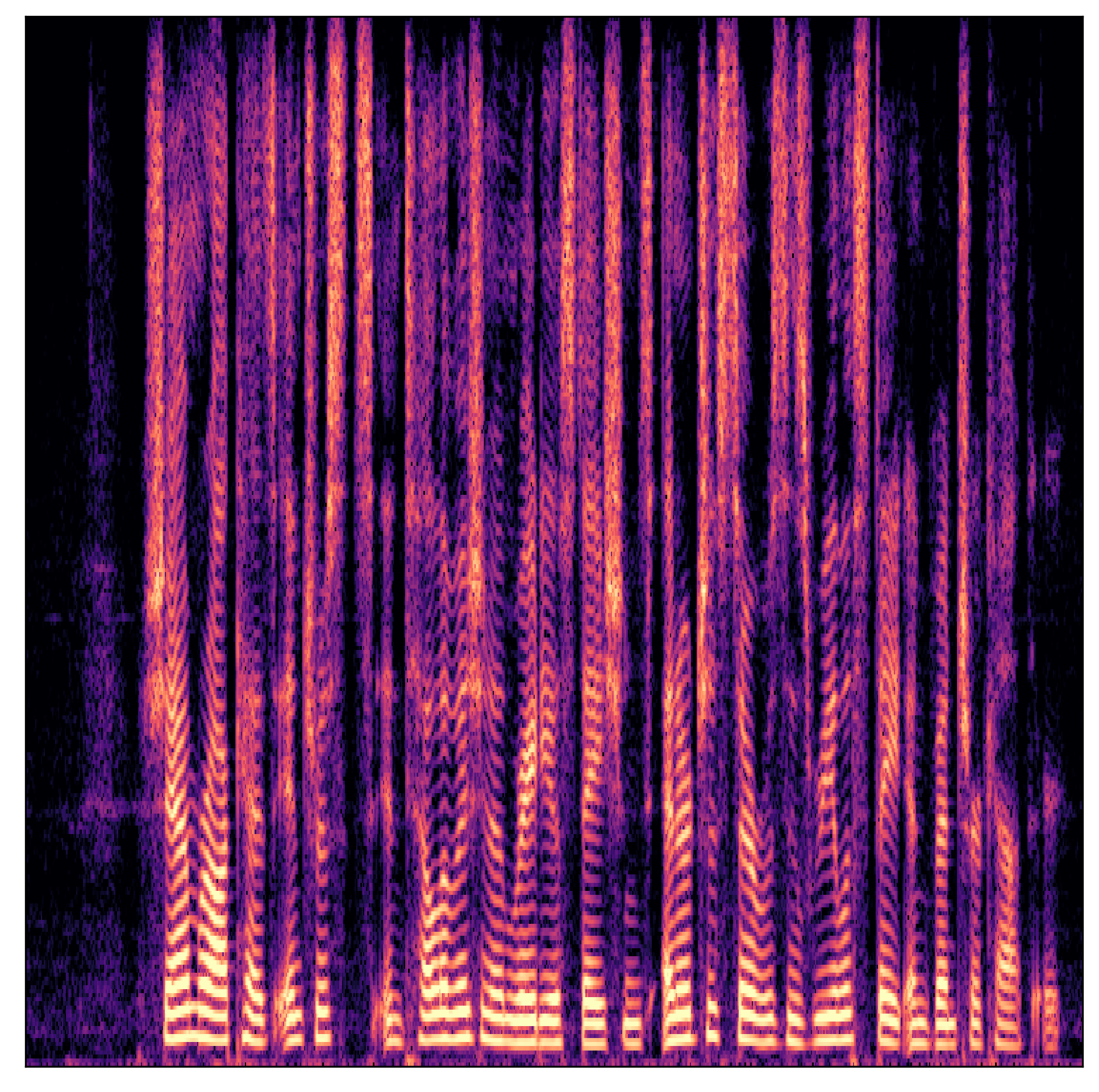};

\end{axis}

 \begin{axis}
[
    at={(regensgmse.south east)},
    xshift = 0.01\textwidth,
    yshift = -0.0195\textwidth,
    width = 0.13\textwidth,
    height = 0.308\textwidth,
    hide axis,
]
\addplot graphics[xmin=0,ymin=0,xmax=1,ymax=1] {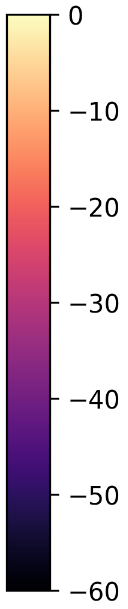};
\end{axis}

\end{tikzpicture}
 \vspace{-2em}
\label{fig:spectrograms-enh}
\end{figure*}

\begin{figure*}

\caption{\protect\centering \textit{Anechoic, reverberant and processed utterances from WS0+Reverb. Input $T_\mathrm{60}$ is 1.06~$\mathrm{s}$.
Formant structure is partly destroyed by SGMSE+M
and severe speech distortions are observed in the NCSN++M sample. StoRM corrects the distortions and reproduces the formant structure without residual reverberation.
}}
\begin{tikzpicture}[scale=0.95, transform shape]
 \protect\centering
\begin{axis}
[
    name={clean},
    title = {Anechoic},
    axis line style={draw=none},
    xmin = 0, xmax = \xmax,
    ymin = 0, ymax = 8000,
    xtick = {0, 1, 2, 3, 4, 5},
    xticklabel style = {xshift=5pt},
    ytick = {0, 2000, ..., 8000},
    yticklabel style = {yshift=5pt},
    yticklabels={0, 2, 4, 6},
    xlabel = {Time [s]},
    ylabel = {Frequency [kHz]},
    width =\spectrow,
    height =\spectrow
]
\addplot graphics[xmin=0,ymin=0,xmax=\xmax,ymax=8000] {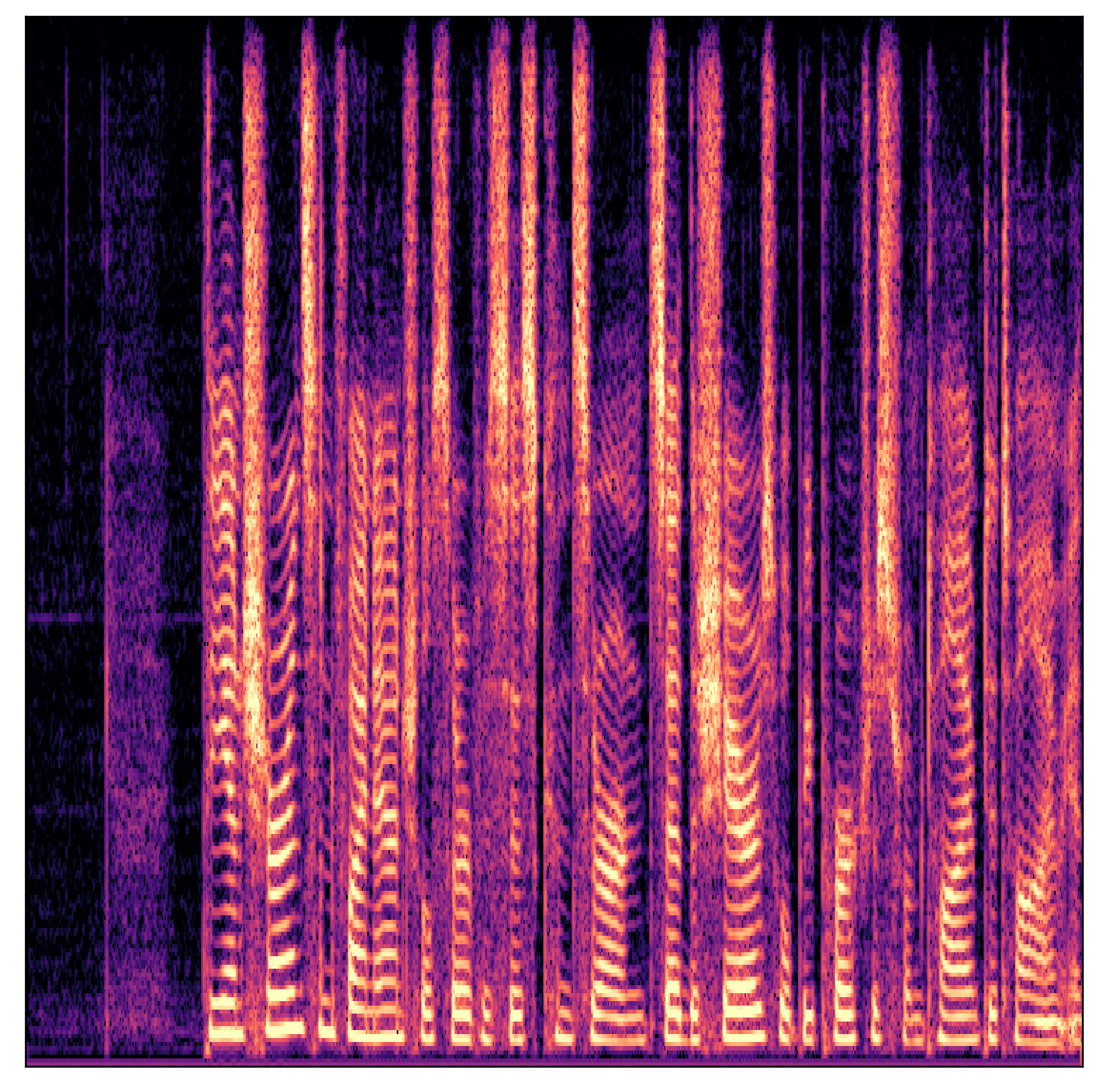};
\end{axis}

\begin{axis}
[
    name = {noisy},
    title = {Reverberant},
    at={(clean.south east)},
    xshift = \spectroxs,
    axis line style={draw=none},
    xmin = 0, xmax = \xmax,
    ymin = 0, ymax = 8000,
    xtick = {0, 1, 2, 3, 4, 5},
    xticklabel style = {xshift=5pt},
    yticklabels=\empty,
    xlabel = {Time [s]},
    width =\spectrow,
    height =\spectrow,
]
\addplot graphics[xmin=0,ymin=0,xmax=\xmax,ymax=8000] {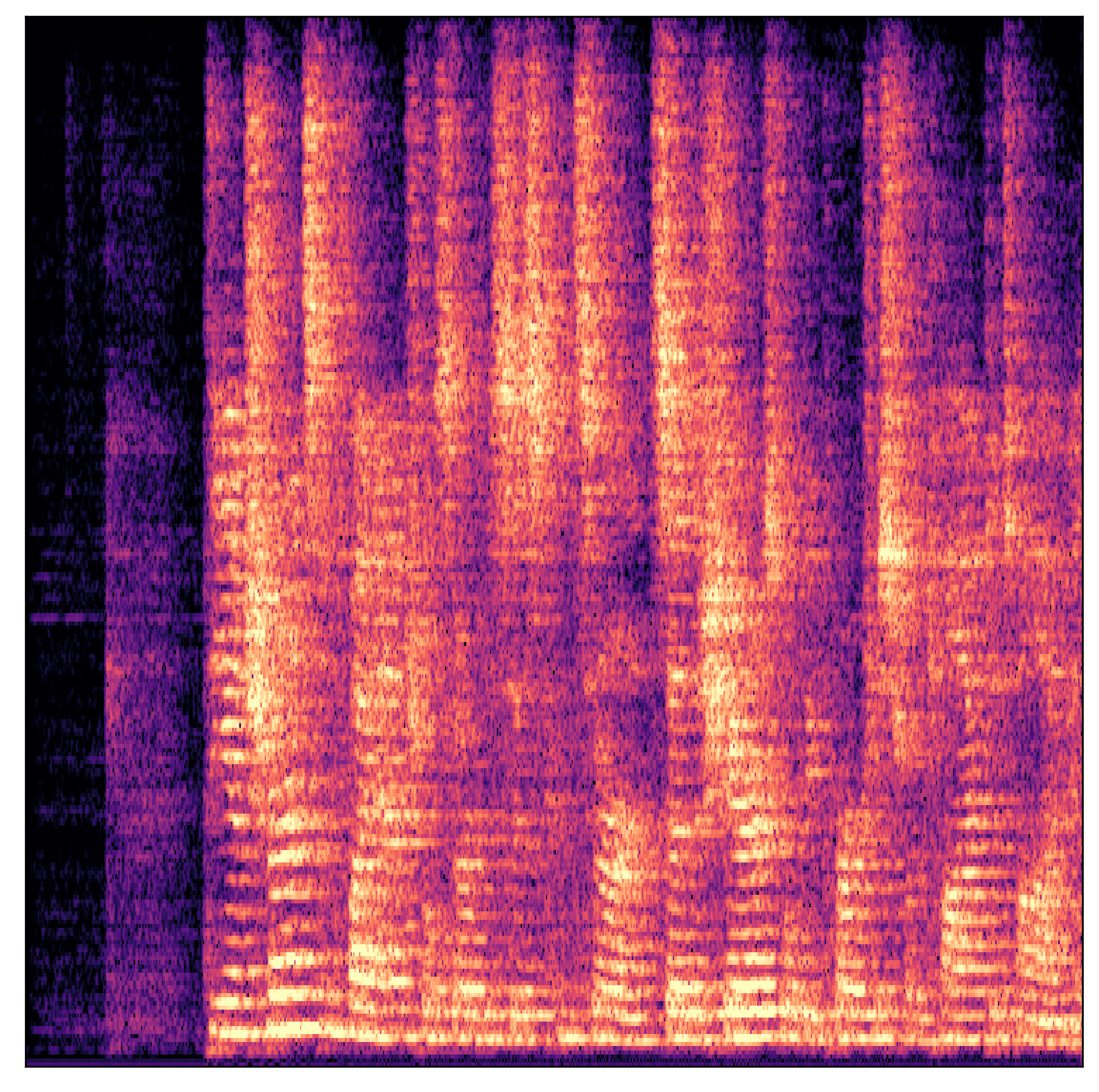};
\end{axis}

\begin{axis}
[
    name = {ncsn},
    title = {NCSN++M},
    at={(noisy.south east)},
    xshift = \spectroxs,
    axis line style={draw=none},
    xmin = 0, xmax = \xmax,
    ymin = 0, ymax = 8000,
    xtick = {0, 1, 2, 3, 4, 5},
    xticklabel style = {xshift=5pt},
    yticklabels=\empty,
    xlabel = {Time [s]},
    width =\spectrow,
    height =\spectrow,
]
\addplot graphics[xmin=0,ymin=0,xmax=\xmax,ymax=8000] {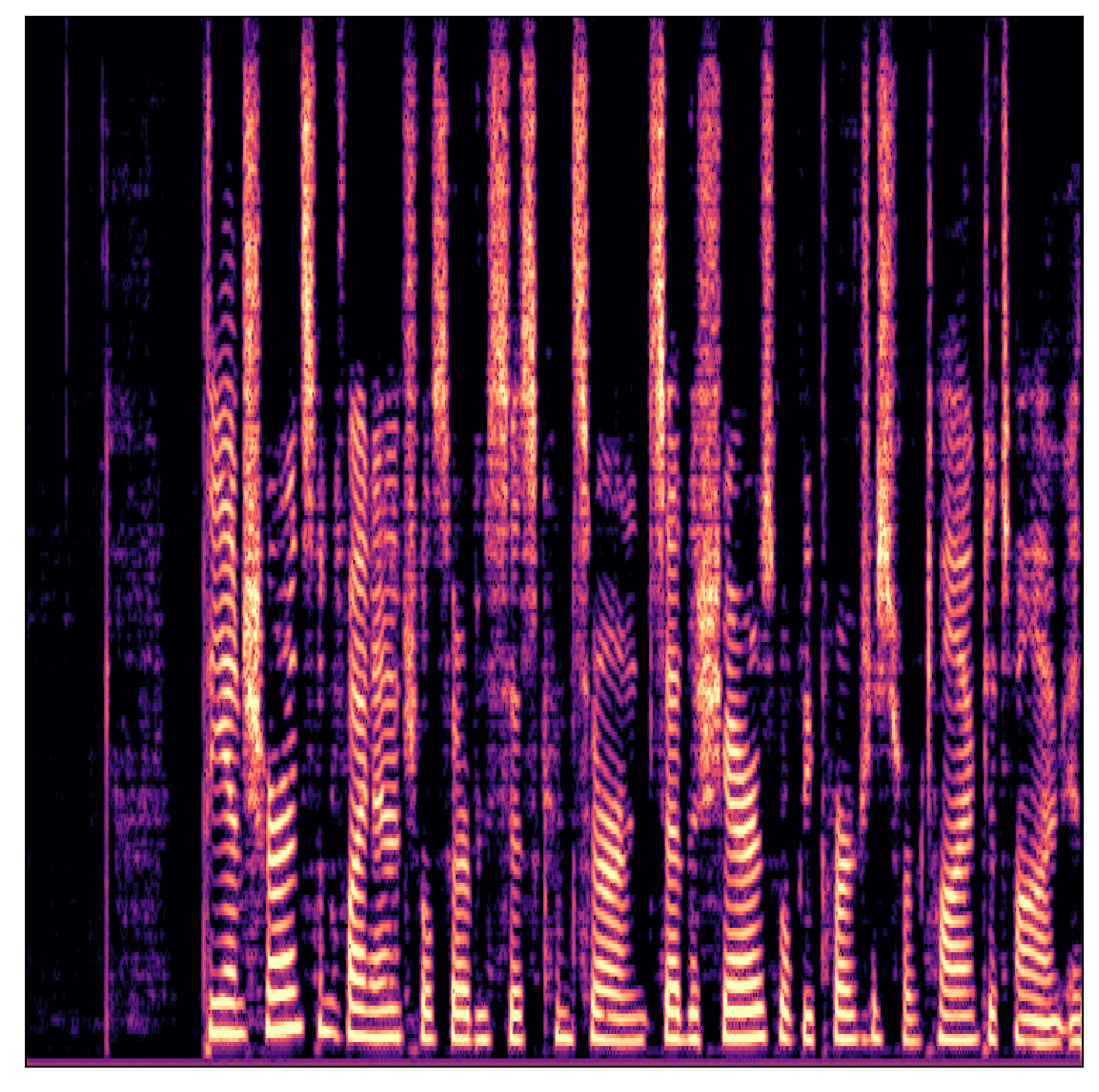};

\end{axis}

\begin{axis}
[
    name = {sgmse},
    title = {SGMSE+M},
    at={(ncsn.south east)},
    xshift = \spectroxs,
    axis line style={draw=none},
    xmin = 0, xmax = \xmax,
    ymin = 0, ymax = 8000,
    xtick = {0, 1, 2, 3, 4, 5},
    xticklabel style = {xshift=5pt},
    yticklabels=\empty,
    xlabel = {Time [s]},
    width =\spectrow,
    height =\spectrow,
]
\addplot graphics[xmin=0,ymin=0,xmax=\xmax,ymax=8000] {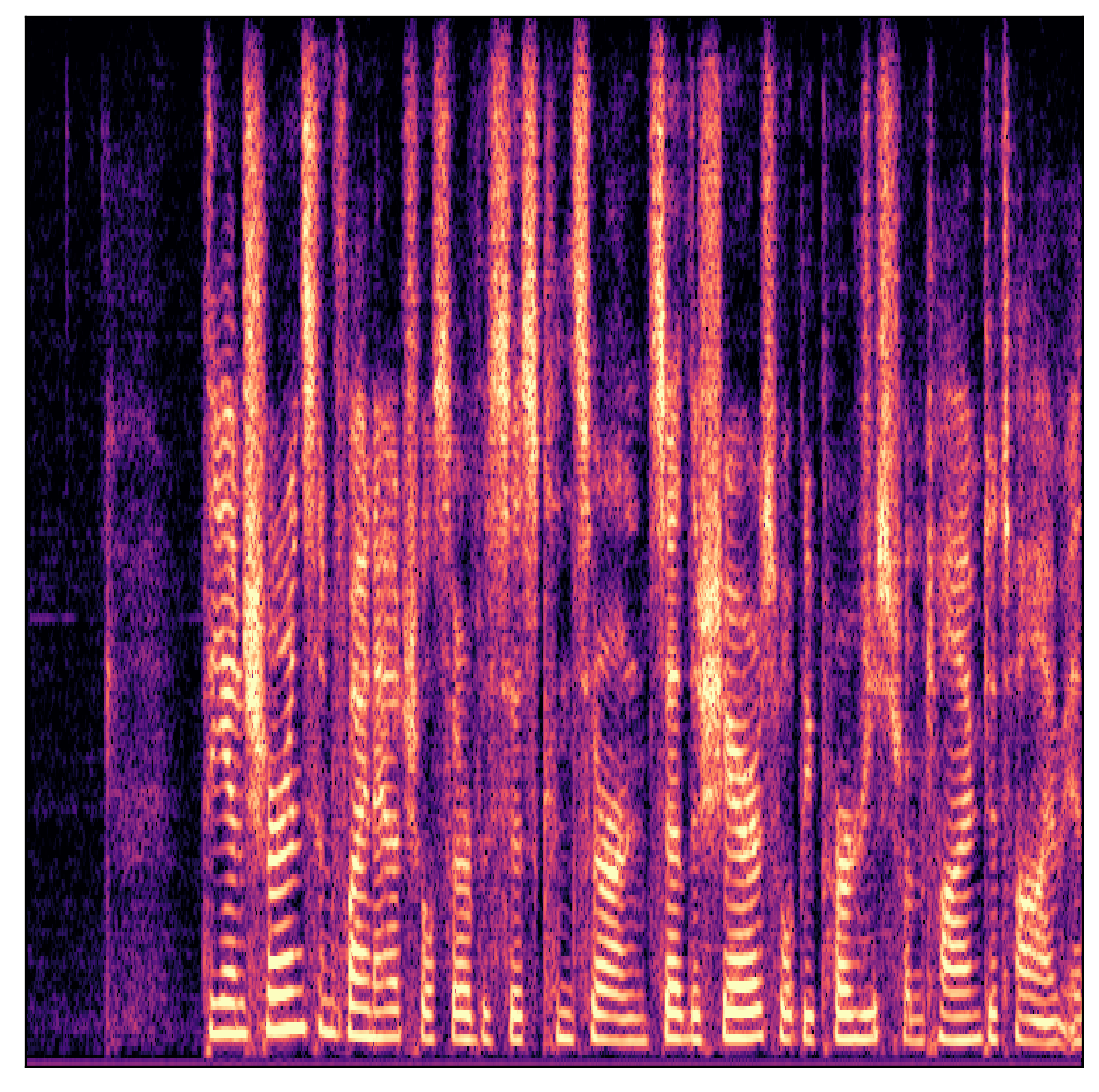};

\end{axis}

\begin{axis}
[
    name = {regensgmse},
    title = {StoRM},
    at={(sgmse.south east)},
    xshift = \spectroxs,
    axis line style={draw=none},
    xmin = 0, xmax = \xmax,
    ymin = 0, ymax = 8000,
    xtick = {0, 1, 2, 3, 4, 5},
    xticklabel style = {xshift=5pt},
    yticklabels=\empty,
    xlabel = {Time [s]},
    width =\spectrow,
    height =\spectrow,
]
\addplot graphics[xmin=0,ymin=0,xmax=\xmax,ymax=8000] {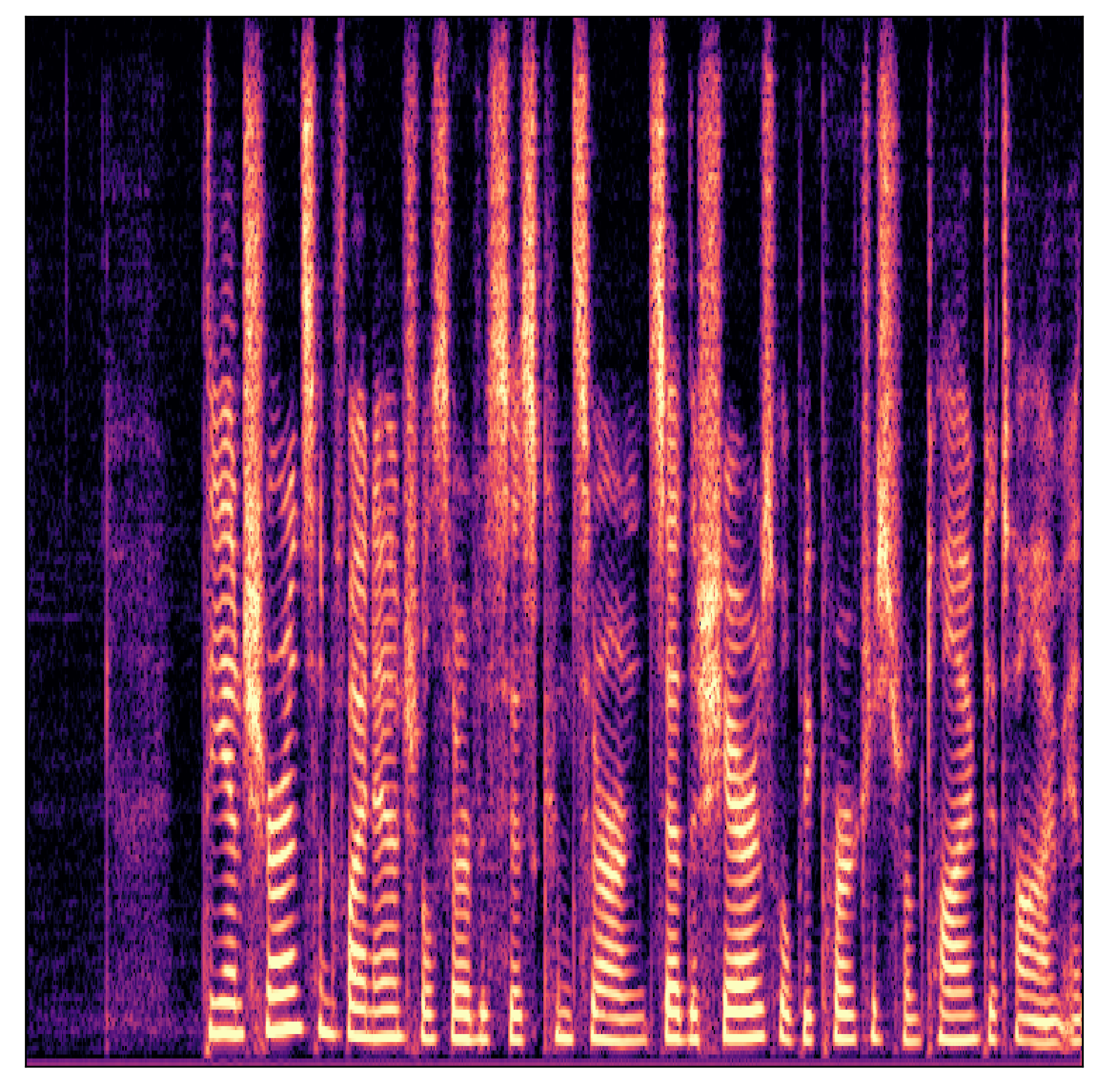};

\end{axis}

 \begin{axis}
[
    at={(regensgmse.south east)},
    xshift = 0.01\textwidth,
    yshift = -0.0195\textwidth,
    width = 0.13\textwidth,
    height = 0.308\textwidth,
    hide axis,
]
\addplot graphics[xmin=0,ymin=0,xmax=1,ymax=1] {graphics/spectros/colorbar.png};
\end{axis}

\end{tikzpicture}
 
 \vspace{-1em}
\label{fig:spectrograms-derev}
\end{figure*}

%% file: tables/mismatch.tex
\begin{table}[]
\caption{\centering\textit{Denoising results on WSJ0+Chime$^\star$ (with input SNRs between 0dB and 20dB) in matched and mismatched settings. In the mismatch setting, the approaches are trained on VoiceBank/DEMAND. All methods use the NCSN++M architecture. SGMSE+M and StoRM use $50$ diffusion steps.
}}
    \hspace{-0.65em}
    \scalebox{0.75}{
    \begin{tabular}{c|c|cccc}
    
\toprule 
Method & Match & WV-MOS & 
PESQ & ESTOI & SI-SDR \\

\midrule
\midrule

\multirow{2}{*}{NCSN++M} 
& \cmark & 3.78 & 
2.73 & \textbf{0.94} & \textbf{20.0} \\ 
& \xmark & 3.33 (-0.35) & 
2.14 (-0.59) & \textbf{0.90} (-0.04) & \textbf{17.6} (-2.4) \\
\midrule

\multirow{2}{*}{SGMSE+M} 
& \cmark & 3.84 & 
2.96 & 0.92 & 17.4 \\
& \xmark & 3.61 (-0.23) & 
\textbf{2.48} (\textbf{-0.48}) & \textbf{0.90} (\textbf{-0.02}) & 15.7 (\textbf{-1.7}) \\
\midrule

\multirow{2}{*}{StoRM} 
& \cmark & \textbf{3.93} & 
\textbf{3.01} & 0.93  & 18.5\\
& \xmark & \textbf{3.71} (\textbf{-0.22}) & 
2.47 (-0.54) & \textbf{0.90} (-0.03) & 16.8 (\textbf{-1.7}) \\

\bottomrule

\end{tabular}
}
\vspace{-1em}
\label{tab:results:mismatch_wsj0-0-20}
\end{table}

%% file: tables/predictors.tex
\begin{table} 
    \caption{\centering\textit{Denoising results on WSJ0+Chime for StoRM using matched and mismatched initial predictors. The predictor architecture used for training is NCSN++M. All approaches use the NCSN++M as score network and $N=50$ steps
    . Values indicate mean and standard deviation.}}
    \centering
    \scalebox{0.80}{
    \begin{tabular}{c|c|ccc}
    
\toprule 
Initial Predictor & Matched & 
PESQ & ESTOI & SI-SDR \\

\midrule
\midrule
Mixture & - & 
1.38 $\pm$ 0.32 & 0.65 $\pm$ 0.18 & $\,$$\,$$\,$4.3 $\pm$ 5.8 \\ \midrule

NCSN++M & \cmark & 
\textbf{2.53 $\pm$ 0.63} & \textbf{0.88 $\pm$ 0.09} & \textbf{14.7 $\pm$ 4.3} \\

GaGNet \cite{Li2022GaGNet} & \xmark & 
2.52 $\pm$ 0.62 & 0.87 $\pm$ 0.09 & \textbf{14.7 $\pm$ 4.1} \\

ConvTasNet \cite{Luo2018} & \xmark & 
2.36 $\pm$ 0.60 & 0.86 $\pm$ 0.09 & $\,$$\,$$\,$9.9 $\pm$ 1.7 \\

\midrule
\bottomrule

\end{tabular}
}
    \label{tab:results:predictors}
\end{table}

%% file: plots/barplot_asr.tex
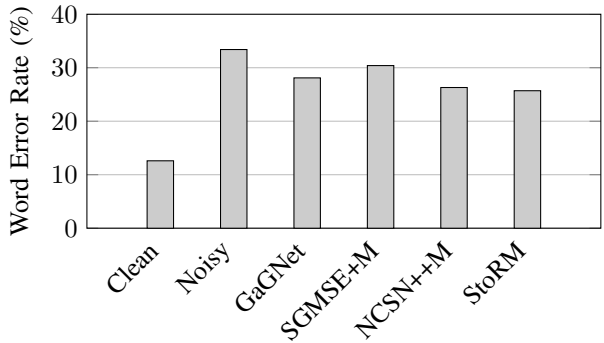
\begin{figure}
    \centering
    \caption{\centering\textit{ASR results for speech enhancement on TIMIT+Chime. 
    Diffusion models use $N=50$ steps.
    }}
    \begin{tikzpicture}
    
\begin{axis}[ybar,
width=0.95\columnwidth, height=0.5\columnwidth,
enlarge x limits=0.2,
xtick=data,
ylabel={Word Error Rate (\%)},
symbolic x coords={Clean,Noisy,GaGNet,SGMSE+M,NCSN++M,StoRM},
x tick label style={rotate=45,anchor=east},
xtick style={draw=none},
ytick = {0, 10, ..., 40},
ymin=0, ymax=40,
ymajorgrids
]
\addplot[draw=black,fill=black!20] table[x=Method,y=WER,col sep=comma] {quartznet_asr.csv};

\end{axis}

    \end{tikzpicture}
    \vspace{-1em}
    \label{fig:results:asr}
\end{figure}

%% file: plots/boxplot_mushra.tex
\begin{figure}
    \caption{\centering \textit{Listening test results. CQS is the "continuous quality scale" on which participants are asked to rate. Inner line represents the median. 9 participants rated 10 samples randomly selected from WSJ0+Chime and WSJ0+Reverb.}}
    \hspace{0.019\textwidth}
    \includegraphics[width=0.85\columnwidth]{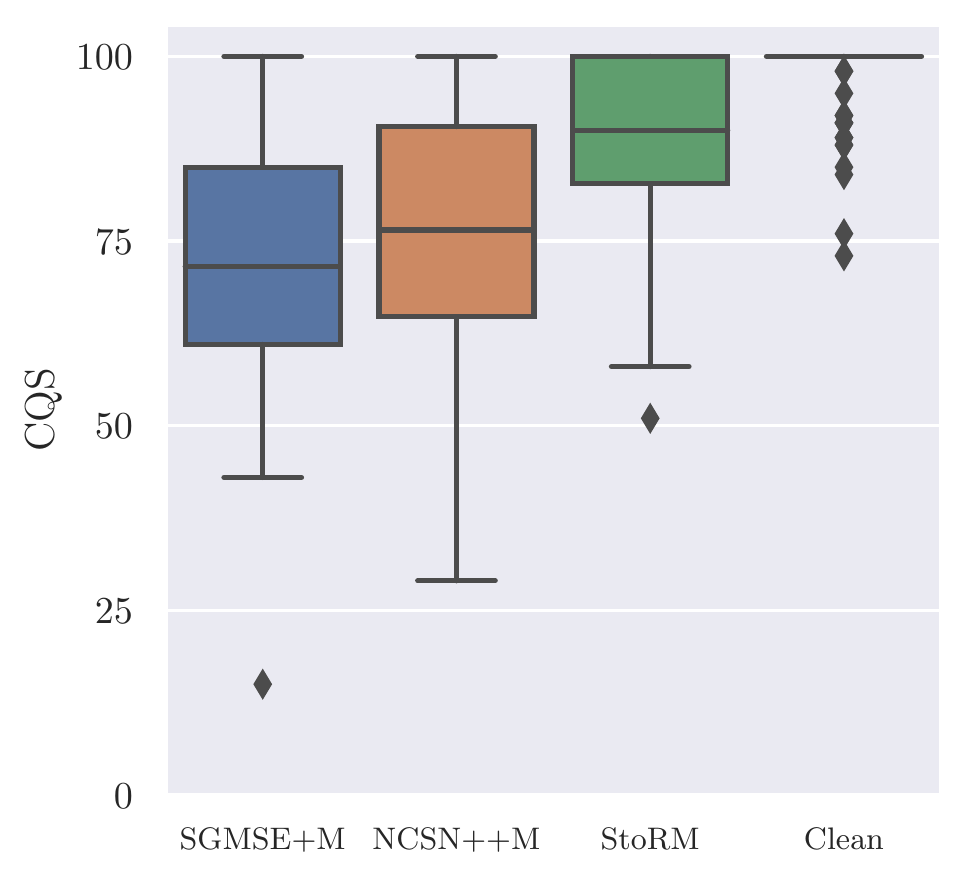}
    \vspace{-0.5em}
    \label{fig:results:mushra}
\end{figure}

%% file: tables/conditioning.tex
\begin{table}[]
\caption{\centering\textit{Denoising results on WSJ0+Chime for StoRM using different conditioning inputs for the score network. Values indicate mean and standard devation. All approaches use NCSN++M as backbone architecture and $N=50$ steps.
}}
    \centering
    \scalebox{1.0}{
    \begin{tabular}{c|ccc}
    
\toprule 
Conditioning & 
PESQ & ESTOI & SI-SDR \\

\midrule
\midrule

Noisy & 
2.30 $\pm$ 0.60 & 0.84 $\pm$ 0.10 & 11.5 $\pm$ 5.2 \\

PostDenoiser & 
2.50 $\pm$ 0.62 & 0.87 $\pm$ 0.09 & 14.7 $\pm$ 4.3 \\
\midrule

Both & 
\textbf{2.53 $\pm$ 0.63} & \textbf{0.88 $\pm$ 0.08} & \textbf{15.1 $\pm$ 4.2} \\

\midrule
\bottomrule

\end{tabular}
}
\vspace{-1em}
\label{tab:results:conditioning}
\end{table}

%% file: tables/training.tex
\begin{table}[]
\caption{\centering\textit{Denoising results on WSJ0+Chime for StoRM using different training strategies for the score network. All approaches use NCSN++M as backbone architecture and $N=50$ steps for reverse diffusion. 
}}
    \centering
    \scalebox{0.8}{
    \begin{tabular}{ccc|ccc}
    
\toprule 
Pre-train $D_\theta$ & Fine-tune $D_\theta$ & Use $\mathcal{J}^{(\mathrm{Sup})}$ & PESQ & ESTOI & SI-SDR \\

\midrule
\midrule
\xmark & \cmark & \cmark &
 \textbf{2.58} &  \textbf{0.88} & \textbf{15.1}  \\
 
\cmark & \xmark & \xmark &
2.53 & \textbf{0.88} & 14.7  \\

\xmark & \cmark & \xmark &
1.11 & 0.62 & $\,$-0.3 \\
\midrule 

\cmark & \cmark & \cmark & 
 \textbf{2.58}  & \textbf{0.88} & \textbf{15.1} \\

\midrule
\bottomrule

\end{tabular}
}
\vspace{-1em}
\label{tab:results:training}
\end{table}

%% file: sections/conclusion.tex
We presented a generative stochastic regeneration scheme combining a predictive model as initial predictor and a diffusion-based generative approach regenerating the target cues distorted by the first stage. 
On the one hand, the approach improves sample quality compared to pure predictive approaches as it leverages generative modelling to output samples that have high probability on the target posterior distribution manifold, rather than regressing to their mean. On the other hand, it uses predictive power to provide a good initial prediction of the target sample, which avoids typical generative artifacts such as vocalizing and breathing effects, and increases the interference removal performance, especially  in difficult environments.
Intrusive and reference-free instrumental metrics as well as formal listening tests confirmed the superiority of the stochastic regeneration approach over the baselines.
The resulting approach allows efficient sampling, requiring fewer steps and avoiding the use of Annealed Langevin Dynamics correction during reverse diffusion, thus reducing computational complexity by an order of magnitude without sacrificing quality, compared to the original diffusion model.